\documentclass[12pt,a4paper]{article}
\pdfoutput=1
\usepackage{bigstrut}
\usepackage{multirow}
\usepackage{lineno}  % for line numbering during review
\usepackage{lscape}
\usepackage{graphicx}  % to include figures (can also use other packages)
\usepackage{xspace} % To avoid problems with missing or double spaces after
\usepackage{color}
\usepackage{colortbl}
\usepackage{amsmath} % Adds a large collection of math symbols
\usepackage{ifthen} % for conditional statements
\graphicspath{{./figs/}} % Make Latex search fig subdir for figures

\newboolean{pdflatex}
\setboolean{pdflatex}{true} % use this if using non-eps figures
\newboolean{articletitles}
\setboolean{articletitles}{true} % False removes titles in references

\newboolean{uprightparticles}
\setboolean{uprightparticles}{false} %Set to true to get roman particle symbols

\usepackage{amssymb}
\usepackage{amsfonts}
\usepackage{upgreek} % Adds in support for greek letters in roman typeset

\usepackage{hyperref}    % Hyperlinks in references
\usepackage[all]{hypcap} % Internal hyperlinks to floats.

\def\lhcb {LHCb\xspace}
\def\ux85 {UX85\xspace}

\ifthenelse{\boolean{uprightparticles}}%
{

 \def\PDelta      {\ensuremath{\Delta}\xspace}                 
 \def\PXi      {\ensuremath{\Xi}\xspace}                 
 \def\PLambda      {\ensuremath{\Lambda}\xspace}                 
 \def\PSigma      {\ensuremath{\Sigma}\xspace}                 
 \def\POmega      {\ensuremath{\Omega}\xspace}                 
 \def\PUpsilon      {\ensuremath{\Upsilon}\xspace}                 
 
 \def\PB      {\ensuremath{\mathrm{B}}\xspace}                 
                  
 \def\PD      {\ensuremath{\mathrm{D}}\xspace}

 \def\PK      {\ensuremath{\mathrm{K}}\xspace}

 \def\Pb      {\ensuremath{\mathrm{b}}\xspace}                 
 \def\Pc      {\ensuremath{\mathrm{c}}\xspace}

 \def\Pi      {\ensuremath{\mathrm{i}}\xspace}

}
{

 \mathchardef\PDelta="7101
 \mathchardef\PXi="7104
 \mathchardef\PLambda="7103
 \mathchardef\PSigma="7106
 \mathchardef\POmega="710A
 \mathchardef\PUpsilon="7107
                  
 \def\PB      {\ensuremath{B}\xspace}                 
                  
 \def\PD      {\ensuremath{D}\xspace}

 \def\PK      {\ensuremath{K}\xspace}

 \def\Pb      {\ensuremath{b}\xspace}                 
 \def\Pc      {\ensuremath{c}\xspace}

 \def\Pi      {\ensuremath{i}\xspace}

}

\def\cquark    {\ensuremath{\Pc}\xspace}

\def\bquark    {\ensuremath{\Pb}\xspace}

\def\kaon  {\ensuremath{\PK}\xspace}
  \def\Kbar  {\kern 0.2em\overline{\kern -0.2em \PK}{}\xspace}

\def\Kz    {\ensuremath{\kaon^0}\xspace}
\def\Kzb   {\ensuremath{\Kbar^0}\xspace}
\def\KzKzb {\ensuremath{\Kz \kern -0.16em \Kzb}\xspace}
\def\Kp    {\ensuremath{\kaon^+}\xspace}
\def\Km    {\ensuremath{\kaon^-}\xspace}

\def\KpKm  {\ensuremath{\Kp \kern -0.16em \Km}\xspace}

  \def\Dbar    {\kern 0.2em\overline{\kern -0.2em \PD}{}\xspace}
\def\D       {\ensuremath{\PD}\xspace}

\def\Dz      {\ensuremath{\D^0}\xspace}
\def\Dzb     {\ensuremath{\Dbar^0}\xspace}
\def\DzDzb   {\ensuremath{\Dz {\kern -0.16em \Dzb}}\xspace}
\def\Dp      {\ensuremath{\D^+}\xspace}
\def\Dm      {\ensuremath{\D^-}\xspace}

\def\DpDm    {\ensuremath{\Dp {\kern -0.16em \Dm}}\xspace}

  \def\Bbar    {\kern 0.18em\overline{\kern -0.18em \PB}{}\xspace}

  \def\Y#1S{\ensuremath{\PUpsilon{(#1S)}}\xspace}% no space before {...}!

\def\Lbar {\ensuremath{\kern 0.1em\overline{\kern -0.1em\Lambda\kern -0.05em}\kern 0.05em{}}\xspace}

\def\to                 {\ensuremath{\rightarrow}\xspace}

\def\CP                {\ensuremath{C\!P}\xspace}

\def\AT#1     {\ensuremath{A_{\mathrm{T}}^{#1}}\xspace}           % 2

\def\C#1      {\ensuremath{\mathcal{C}_{#1}}\xspace}                       % 9
\def\Cp#1     {\ensuremath{\mathcal{C}_{#1}^{'}}\xspace}                    % 7
\def\Ceff#1   {\ensuremath{\mathcal{C}_{#1}^{\mathrm{(eff)}}}\xspace}        % 9  
\def\Cpeff#1  {\ensuremath{\mathcal{C}_{#1}^{'\mathrm{(eff)}}}\xspace}       % 7
\def\Ope#1    {\ensuremath{\mathcal{O}_{#1}}\xspace}                       % 2
\def\Opep#1   {\ensuremath{\mathcal{O}_{#1}^{'}}\xspace}                    % 7

\newcommand{\tev}{\ensuremath{\mathrm{\,Te\kern -0.1em V}}\xspace}
\newcommand{\gev}{\ensuremath{\mathrm{\,Ge\kern -0.1em V}}\xspace}
\newcommand{\mev}{\ensuremath{\mathrm{\,Me\kern -0.1em V}}\xspace}
\newcommand{\kev}{\ensuremath{\mathrm{\,ke\kern -0.1em V}}\xspace}
\newcommand{\ev}{\ensuremath{\mathrm{\,e\kern -0.1em V}}\xspace}
\newcommand{\gevc}{\ensuremath{{\mathrm{\,Ge\kern -0.1em V\!/}c}}\xspace}
\newcommand{\mevc}{\ensuremath{{\mathrm{\,Me\kern -0.1em V\!/}c}}\xspace}
\newcommand{\gevcc}{\ensuremath{{\mathrm{\,Ge\kern -0.1em V\!/}c^2}}\xspace}
\newcommand{\gevgevcccc}{\ensuremath{{\mathrm{\,Ge\kern -0.1em V^2\!/}c^4}}\xspace}
\newcommand{\mevcc}{\ensuremath{{\mathrm{\,Me\kern -0.1em V\!/}c^2}}\xspace}

\def\mum  {\ensuremath{\,\upmu\rm m}\xspace}

\newcommand{\chisq}{\ensuremath{\chi^2}\xspace}

\def\gsim{{~\raise.15em\hbox{$>$}\kern-.85em
          \lower.35em\hbox{$\sim$}~}\xspace}
\def\lsim{{~\raise.15em\hbox{$<$}\kern-.85em
          \lower.35em\hbox{$\sim$}~}\xspace}

\def\evtgen     {\mbox{\textsc{EvtGen}}\xspace}
\def\pythia     {\mbox{\textsc{Pythia}}\xspace}

\def\geant      {\mbox{\textsc{Geant4}}\xspace}

\def\photos     {\mbox{\textsc{Photos}}\xspace}

\def\tell1  {TELL1\xspace}
\def\ukl1   {UKL1\xspace}

\usepackage{cite} % Allows for ranges in citations
\usepackage{mciteplus}

\begin{document}

%%%%%%%%%%%%%%%%%%%%%%%%%
%%%%% Title     %%%%%%%%%
%%%%%%%%%%%%%%%%%%%%%%%%%
\renewcommand{\thefootnote}{\fnsymbol{footnote}}
\setcounter{footnote}{1}

% %%%%%%% CHOOSE --------
% \input{title-LHCb-ANA}
% \input{title-LHCb-CONF}
% $Id: title-LHCb-PAPER.tex 16065 2012-02-20 22:49:03Z uegede $
% ===============================================================================
% Purpose: LHCb-PAPER journal paper title page template
% Author: 
% Created on: 2010-09-25
% ===============================================================================

%%%%%%%%%%%%%%%%%%%%%%%%%
%%%%%  TITLE PAGE  %%%%%%
%%%%%%%%%%%%%%%%%%%%%%%%%
\begin{titlepage}
\pagenumbering{roman}

% Header ---------------------------------------------------
\vspace*{-1.5cm}
\centerline{\large EUROPEAN ORGANIZATION FOR NUCLEAR RESEARCH (CERN)}
\vspace*{1.0cm}
\hspace*{-0.5cm}
\begin{tabular*}{\linewidth}{lc@{\extracolsep{\fill}}r}
\ifthenelse{\boolean{pdflatex}}% Logo format choice
{\vspace*{-3.0cm}\mbox{\!\!\!\includegraphics[width=.14\textwidth]{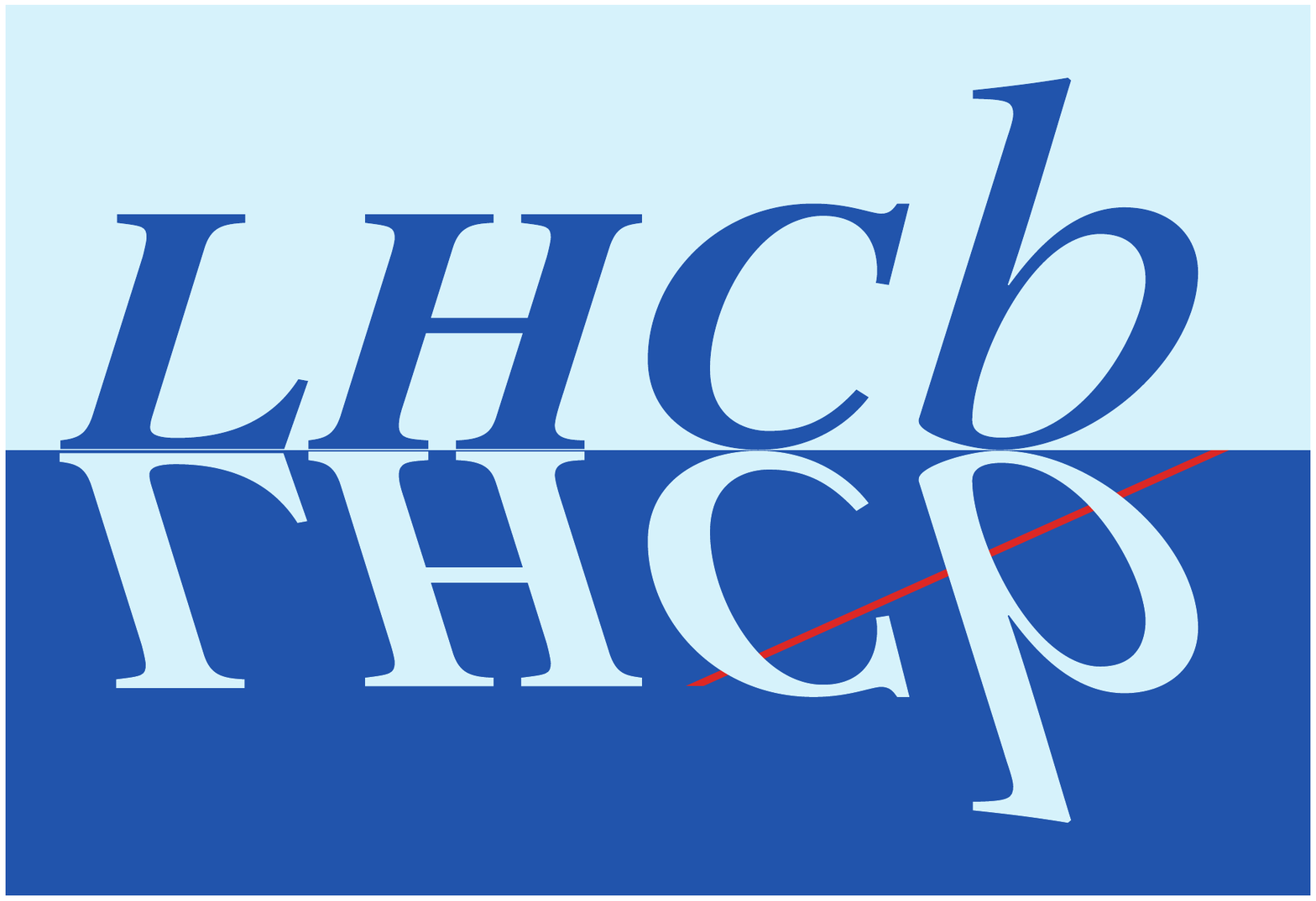}} & &}%
{\vspace*{-1.2cm}\mbox{\!\!\!\includegraphics[width=.12\textwidth]{lhcb-logo.eps}} & &}%
\\
 & & CERN-PH-EP-2012-161 \\  % ID 
 & & LHCb-PAPER-2012-002 \\  % ID 
 & & October 23, 2012 \\ %\today \\ % Date - Can also hardwire e.g.: 23 March 2010
 & & \\
% not in paper \hline
\end{tabular*}

\vspace*{0.5cm}

% Title --------------------------------------------------
{\bf\boldmath\huge
\begin{center}
Measurement of $b$-hadron branching fractions for two-body decays into charmless charged hadrons
\end{center}
}

\vspace*{0cm}

% Authors -------------------------------------------------
\begin{center}
The LHCb collaboration\footnote{Authors are listed on the following pages.}
\end{center}

\vspace{0cm}

% Abstract -----------------------------------------------
\begin{abstract}
  \noindent
Based on data corresponding to an integrated luminosity of 0.37 $\mathrm{fb}^{-1}$ collected by the LHCb experiment in 2011, the following ratios of branching fractions are measured:
\begin{eqnarray*}
\mathcal{B}\left(B^{0}\rightarrow\pi^{+}\pi^{-}\right) /\,\mathcal{B}\left(B^{0}\rightarrow K^+\pi^-\right) & = & 0.262\pm 0.009\pm 0.017,\\
(f_{s} / f_{d}) \cdot \mathcal{B}\left(B^{0}_{s}\rightarrow K^{+}K^{-}\right) /\, \mathcal{B}\left(B^{0}\rightarrow K^+\pi^-\right) & = & 0.316\pm 0.009\pm 0.019,\\
(f_{s} / f_{d}) \cdot \mathcal{B}\left(B^0_{s}\rightarrow\pi^+ K^-\right) /\, \mathcal{B}\left(B^{0}\rightarrow K^+\pi^-\right) & = & 0.074 \pm 0.006\pm 0.006,\\
(f_{d} / f_{s}) \cdot \mathcal{B}\left(B^{0} \rightarrow K^{+}K^{-}\right) /\, \mathcal{B}\left(B^{0}_s\rightarrow K^+K^-\right) & = & 0.018 \,^{+\,0.008}_{-\,0.007} \pm 0.009,\\
(f_{s} / f_{d}) \cdot \mathcal{B}\left(B^{0}_{s}\rightarrow \pi^{+}\pi^{-}\right) /\, \mathcal{B}\left(B^{0}\rightarrow \pi^+\pi^-\right) & = & 0.050 \,^{+\,0.011}_{-\,0.009} \pm 0.004,\\
\mathcal{B}\left(\Lambda^0_b\rightarrow p\pi^-\right) /\, \mathcal{B}\left(\Lambda^0_b\rightarrow pK^-\right) & = & 0.86 \pm 0.08\pm 0.05,
\end{eqnarray*} 
where the first uncertainties are statistical and the second systematic.
Using the current world average of $\mathcal{B}\left(B^{0}\rightarrow K^+\pi^-\right)$ and the ratio of the strange to light neutral $B$ meson production $f_s/f_d$ measured by LHCb, we obtain: 
\begin{eqnarray*}
\mathcal{B}\left(B^{0}\rightarrow\pi^{+}\pi^{-}\right) & = & (5.08 \pm 0.17 \pm 0.37)\times 10^{-6},\\
\mathcal{B}\left(B^{0}_{s}\rightarrow K^{+}K^{-}\right) & = & (23.0 \pm 0.7 \pm  2.3)\times 10^{-6},\\
\mathcal{B}\left(B^0_{s}\rightarrow\pi^+ K^-\right) & = & (5.4 \pm 0.4 \pm  0.6)\times 10^{-6},\\
\mathcal{B}(B^0 \rightarrow K^+K^-) & = & (0.11 \,^{+\,0.05}_{-\,0.04} \pm 0.06)\times10^{-6},\\
\mathcal{B}(B^0_s \rightarrow \pi^+\pi^-) & = & (0.95 \,^{+\,0.21}_{-\,0.17} \pm 0.13)\times10^{-6}.
\end{eqnarray*} 
The measurements of $\mathcal{B}\left(B^{0}_{s}\rightarrow K^{+}K^{-}\right)$, $\mathcal{B}\left(B^0_{s}\rightarrow\pi^+ K^-\right)$ and $\mathcal{B}(B^0 \rightarrow K^+K^-)$ are the most precise to date. The decay mode $B^0_s \rightarrow \pi^+\pi^-$ is observed for the first time with a significance of more than $5\sigma$.
\end{abstract}

\vspace*{0cm}

%\begin{center}
%  Submitted to Journal of High Energy Physics
%\end{center}

\vspace{\fill}

\end{titlepage}

%%%%%%%%%%%%%%%%%%%%%%%%%%%%%%%%
%%%%%  EOD OF TITLE PAGE  %%%%%%
%%%%%%%%%%%%%%%%%%%%%%%%%%%%%%%%

%  empty page follows the title page ----
\newpage
\setcounter{page}{2}
\mbox{~}
\newpage

% Author List ----------------------------
%  You need to get a new author list!
%%%%%%%%%%%%%%%%%%%%%%%%%%%%%%%%%%%%%%%%%%
\centerline{\large\bf LHCb collaboration}
\begin{flushleft}
\small
R.~Aaij$^{38}$, 
C.~Abellan~Beteta$^{33,n}$, 
B.~Adeva$^{34}$, 
M.~Adinolfi$^{43}$, 
C.~Adrover$^{6}$, 
A.~Affolder$^{49}$, 
Z.~Ajaltouni$^{5}$, 
J.~Albrecht$^{35}$, 
F.~Alessio$^{35}$, 
M.~Alexander$^{48}$
% ...

{\footnotesize \it
% ...
$ ^{5}$Clermont Universit\'{e}, Universit\'{e} Blaise Pascal, CNRS/IN2P3, LPC, Clermont-Ferrand, France\\
$ ^{6}$CPPM, Aix-Marseille Universit\'{e}, CNRS/IN2P3, Marseille, France\\
% ...
$ ^{33}$Universitat de Barcelona, Barcelona, Spain\\
$ ^{34}$Universidad de Santiago de Compostela, Santiago de Compostela, Spain\\
$ ^{35}$European Organization for Nuclear Research (CERN), Geneva, Switzerland\\
% ...
$ ^{38}$Nikhef National Institute for Subatomic Physics, Amsterdam, The Netherlands\\
% ...
$ ^{43}$H.H. Wills Physics Laboratory, University of Bristol, Bristol, United Kingdom\\
% ...
$ ^{48}$School of Physics and Astronomy, University of Glasgow, Glasgow, United Kingdom\\
$ ^{49}$Oliver Lodge Laboratory, University of Liverpool, Liverpool, United Kingdom\\
% ...
$ ^{n}$LIFAELS, La Salle, Universitat Ramon Llull, Barcelona, Spain\\
% ...
}
\end{flushleft}
%%%%%%%%%%%%%%%%%%%%%%%%%%%%%%%%%%%%%%%%%%

\cleardoublepage

% %%%%%%%%%%%%% ---------

\renewcommand{\thefootnote}{\arabic{footnote}}
\setcounter{footnote}{0}

%%%%%%%%%%%%%%%%%%%%%%%%%%%%%%%%
%%%%%  Table of Content   %%%%%%
%%%%%%%%%%%%%%%%%%%%%%%%%%%%%%%%
%%%% Uncomment next 2 lines if desired
%\tableofcontents
%\cleardoublepage

%%%%%%%%%%%%%%%%%%%%%%%%%
%%%%% Main text %%%%%%%%%
%%%%%%%%%%%%%%%%%%%%%%%%%

\pagestyle{plain} % restore page numbers for the main text
\setcounter{page}{1}
\pagenumbering{arabic}

% %%%%%%% CHOOSE --------
%% ----------------------------------
%% Line numbering on the left margin 
%% ----------------------------------
%% Uncomment during review phase. 
%% Comment it out before a final submission.
%\linenumbers
%% --------------------------------
% %%%%%%%%%%%%% ---------

% You can include short sections directly in the main tex file.
% However, for larger papers it is desirable to split the text into
% several semiautonomous files, which can be revised independently.
% This is especially useful when developing a document in
% collaboration with several people, since then different parts can be
% edited independently.  This type of file organization is shown here.
% 

\section{Introduction}
\label{sec:intro}

In the quest for physics beyond the Standard Model (SM) in the flavour sector, the study of charmless $H_b \rightarrow h^+h^{\prime -}$ decays, where $H_b$ is a $b$-flavoured meson or baryon, and $h^{(\prime)}$ stands for a pion, kaon or proton, plays an important role.
A simple interpretation of the \CP-violating observables of the charmless two-body $b$-hadron decays in terms of Cabibbo-Kobayashi-Maskawa (CKM) weak phases~\cite{Cabibbo:1963yz,*Kobayashi:1973fv} is not possible. The presence of so-called penguin diagrams in addition to tree diagrams gives non-negligible contributions to the decay amplitude and introduces unknown hadronic factors. This then poses theoretical challenges for an accurate determination of CKM phases. On the other hand, penguin diagrams may have contributions from physics beyond the SM~\cite{Fleischer:1999pa,Gronau:2000md,Lipkin:2005pb,Fleischer:2007hj,Fleischer:2010ib}. These questions have motivated an experimental programme aimed at the measurement of the properties of these decays~\cite{Aubert:2008sb,Belle:2008zza,Aaltonen:2011qt,LHCb-PAPER-2011-029,Aaltonen:2011jv}.

Using data corresponding to an integrated luminosity of $0.37$~$\mathrm{fb}^{-1}$ collected by the LHCb experiment in 2011, we report measurements of the branching fractions $\mathcal{B}$ of the $B^0\rightarrow\pi^+\pi^-$, $B_s^0\rightarrow K^+ K^-$, $B_s^0\rightarrow\pi^+K^-$, $B^0\rightarrow K^+K^-$ and $B_s^0\rightarrow\pi^+\pi^-$ decays. Furthermore, we also measure the ratio of the $\Lambda^0_b\rightarrow p\pi^-$ and $\Lambda^0_b\rightarrow p K^-$ branching fractions. The inclusion of charge-conjugate decay modes is implied throughout the paper.

The ratio of branching fractions between any two of these decays can be expressed as 
\begin{equation}\label{eq:brformula}
\frac{\mathcal{B}(H_b\rightarrow F)}{\mathcal{B}(H^{\prime}_b\rightarrow F^{\prime})}=
\frac{f_{H^{\prime}_b}}{f_{H_b}}\cdot
\frac{N(H_b\rightarrow F)}{N(H^{\prime}_b\rightarrow F^{\prime})}\cdot
\frac{\varepsilon_{\rm rec}(H^{\prime}_b\rightarrow F^{\prime})}{\varepsilon_{\rm rec}(H_b\rightarrow F)}\cdot
\frac{\varepsilon_{\rm PID}(F^{\prime})}{\varepsilon_{\rm PID}(F)}
\end{equation}
where $f_{H_{b}^{(\prime)}}$ is the probability for a $b$ quark to hadronize into a $H_{b}^{(\prime)}$ hadron, $N$ is the observed yield of the given decay to the final state $F ^{(\prime)}$, $\varepsilon_{\rm rec}$ is the overall reconstruction efficiency, excluding particle identification (PID), and $\varepsilon_{\rm  PID}$ is the PID efficiency for the corresponding final state hypothesis. We choose to measure ratios where a better cancellation of systematic uncertainties can be achieved.

\section{Detector, trigger and event selection}
\label{sec:Detector}
The \lhcb detector~\cite{Alves:2008zz} is a single-arm forward spectrometer covering the pseudorapidity range $2<\eta <5$, designed
for the study of particles containing \bquark or \cquark quarks. The detector includes a high-precision tracking system consisting of a
silicon-strip vertex detector surrounding the $pp$ interaction region, a large-area silicon-strip detector located upstream of a dipole
magnet with a bending power of about $4{\rm\,Tm}$, and three stations of silicon-strip detectors and straw drift-tubes placed
downstream. The combined tracking system has momentum resolution $\Delta p/p$ that varies from 0.4\% at 5\gevc to 0.6\% at 100\gevc,
and impact parameter resolution of 20\mum for tracks with high transverse momenta. Charged hadrons are identified using two
ring-imaging Cherenkov (RICH) detectors. Photon, electron and hadron candidates are identified by a calorimeter system consisting of
scintillating-pad and pre-shower detectors, an electromagnetic calorimeter and a hadronic calorimeter. Muons are identified by a muon
system composed of alternating layers of iron and multiwire proportional chambers. The trigger consists of a hardware stage, based
on information from the calorimeter and muon systems, followed by a software stage which performs a full event reconstruction.

The software trigger requires a two-, three- or four-track secondary vertex with a high sum of the transverse momenta of
the tracks, significant displacement from the primary interaction, and at least one track with a transverse momentum exceeding $1.7$~\gevc.
Furthermore, it exploits the impact parameter, defined as the smallest distance between the reconstructed trajectory of the particle and the $pp$ collision vertex, requiring its \chisq to be greater than 16.
A multivariate algorithm is used
for the identification of the secondary vertices~\cite{LHCb-PUB-2011-016}. In addition, a dedicated two-body software trigger is used. To discriminate between signal and background events, this trigger selection imposes requirements on: the quality of the online-reconstructed tracks ($\chi^2$/ndf, where ndf is the number of degrees of freedom), their transverse momenta ($p_\mathrm{T}$) and their impact parameters ($d_\mathrm{IP}$); the distance of closest approach of the daughter particles ($d_\mathrm{CA}$); the transverse momentum of the $b$-hadron candidate ($p_\mathrm{T}^B$), its impact parameter ($d_\mathrm{IP}^B$) and its decay time ($t_{\pi\pi}$, calculated assuming decay into $\pi^+\pi^-$). Only $b$-hadron candidates within the $\pi^+\pi^-$ invariant mass range 4.7--5.9~\gevcc are accepted. The $\pi^+\pi^-$ mass hypothesis is chosen to ensure all charmless two-body $b$-hadron decays are selected using the same criteria. 

The events passing the trigger requirements are then filtered to further reduce the size of the data sample. In addition to tighter requirements on the kinematic variables already used in the software trigger, requirements on the larger of the transverse momenta ($p_\mathrm{T}^h$) and of the impact parameters ($d_\mathrm{IP}^h$) of the daughter particles are applied.
As the rates of the various signals under study span two orders of magnitude, for efficient discrimination against combinatorial background three different sets of kinematic requirements are used to select events for: (A) the measurements of $\mathcal{B}\left(B^{0}\rightarrow\pi^{+}\pi^{-}\right) /\,\mathcal{B}\left(B^{0}\rightarrow K^+\pi^-\right)$, $\mathcal{B}\left(B^{0}_{s}\rightarrow K^{+}K^{-}\right) /\, \mathcal{B}\left(B^{0}\rightarrow K^+\pi^-\right)$ and 
$\mathcal{B}(\Lambda^0_b\rightarrow p K^-) /\, \mathcal{B}(\Lambda^0_b\rightarrow p\pi^-)$; (B)~the measurement of
$\mathcal{B}\left(B^0_{s}\rightarrow\pi^+ K^-\right) /\, \mathcal{B}\left(B^{0}\rightarrow K^+\pi^-\right)$;
(C) the measurements of $\mathcal{B}\left(B^{0} \rightarrow K^{+}K^{-}\right) /\, \mathcal{B}\left(B^{0}_s\rightarrow K^+K^-\right)$ and $\mathcal{B}\left(B^{0}_{s}\rightarrow \pi^{+}\pi^{-}\right) /\, \mathcal{B}\left(B^{0}\rightarrow \pi^+\pi^-\right)$.~The kinematic requirements adopted in each selection are summarized in Table~\ref{tab:selectioncuts}. 

\begin{table}
\caption{Summary of criteria adopted in the event selections A, B and C defined in the text. 
}
\begin{centering}
\begin{tabular}{c|r@{\hspace{-0.7cm}}c|r@{\hspace{-0.7cm}}c|r@{\hspace{-0.7cm}}c}
Variable & \multicolumn{2}{c|}{Selection A} & \multicolumn{2}{c|}{Selection B} & \multicolumn{2}{c}{Selection C}  \\
\hline
Track $p_\mathrm{T}\,[\gevc]$ & \hspace{0.3cm}$>$ & $1.1$ & \hspace{0.3cm}$>$ &$1.2$ & \hspace{0.3cm}$>$ & $1.2$   \\
Track $d_\mathrm{IP}\,[\mathrm{\mu m}]$ & $>$& $150$ & $>$ &$200$ & $>$ &$200$ \\
Track $\chi^2$/ndf & $<$ &$3$ & $<$&$3$ & $<$&$3$ \\
$\mathrm{max}(p_\mathrm{T}^{h^+},\, p_\mathrm{T}^{h^{\prime -}})\,[\gevc]$ & $>$ &$2.8$ & $>$&$3.0$ & $>$&$3.0$ \\
$\mathrm{max}(d_\mathrm{IP}^{h^+},\,d_\mathrm{IP}^{h^{\prime -}})\,[\mathrm{\mu m}]$ & $>$&$300$ & $>$&$400$ & $>$&$400$ \\
$d_{\rm CA}\,[\mathrm{\mu m}]$ & $<$&$80$ & $<$&$80$ & $<$&$80$\\
$d_\mathrm{IP}^B\,[\mathrm{\mu m}]$ & $<$&$60$ & $<$&$60$ & $<$&$60$\\
$p_\mathrm{T}^{B}\,[\gevc]$ & $>$&$2.2$ & $>$&$2.4$ & $>$&$2.8$ \\
$t_{\pi\pi}\,[\textrm{ps}]$ & $>$&$0.9$ & $>$&$1.5$ & $>$&$2.0$ \\
\end{tabular}
\par\end{centering}
\label{tab:selectioncuts}
\end{table}

In order to evaluate the ratios of reconstruction efficiencies $\varepsilon_{\rm rec}$, 
needed to calculate the relative branching fractions of two $H_b\rightarrow h^+h^{\prime -}$ decays, we apply
selection and trigger requirements to fully
simulated events.
The results of this study are summarized in Table~\ref{tab:recratio}, where the uncertainties are due to the finite size of the simulated event samples.
Other sources of systematic uncertainties are negligible at the current level of precision. This is confirmed by studies on samples of $D^0$ mesons decaying into pairs of charged hadrons, where reconstruction efficiencies are determined from data using measured signal yields and current world averages of the corresponding branching fractions.
For the simulation, $pp$ collisions are generated using
\pythia~6.4~\cite{Sjostrand:2006za} with a specific \lhcb
configuration~\cite{LHCb-PROC-2010-056}.  Decays of hadrons
are described by \evtgen~\cite{Lange:2001uf} in which final state
radiation is generated using \photos~\cite{Golonka:2005pn}. The
interaction of the generated particles with the detector and its
response are implemented using the \geant
toolkit~\cite{Allison:2006ve, *Agostinelli:2002hh} as described in
Ref.~\cite{LHCb-PROC-2011-006}.

\begin{table}[t]
    \caption{Ratios of reconstruction efficiencies of the various channels, as determined from Monte Carlo simulation, corresponding to the three event selections of Table~\ref{tab:selectioncuts}. PID efficiencies are not included here. The tight requirement on $t_{\pi\pi}$ used in selection C leads to a sizable difference from unity of the ratios in the last two rows, as the $B_s^0\rightarrow\pi^+\pi^-$ and $B_s^0\rightarrow K^+K^-$ decays proceed mainly via the short lifetime component of the $B^0_s$ meson.}\label{tab:recratio}
  \begin{center}
    \begin{tabular}{c|r@{\hspace{0.1cm}}c@{\hspace{0.1cm}}l|c}
     Selection & \multicolumn{3}{c|}{Efficiency ratio} & Value \\
      \hline 
     \multirow{3}{*}{A} & $\varepsilon_{\rm rec}(B^0\rightarrow K^+\pi^-)$&$/$&$\varepsilon_{\rm rec}(B^0\rightarrow\pi^+\pi^-)$ \bigstrut & $0.98 \pm 0.02$ \\
     & $\varepsilon_{\rm rec}(B^0\rightarrow K^+\pi^-) $&$/$&$\varepsilon_{\rm rec}(B_s^0\rightarrow K^+ K^-)$  & $1.00 \pm 0.02$ \\
    & $\varepsilon_{\rm rec}(\Lambda^0_b\rightarrow p K^-) $&$/$&$\varepsilon_{\rm rec}(\Lambda^0_b\rightarrow p\pi^-)$ & $1.00 \pm 0.02$ \\
      \hline
     B \bigstrut & $\varepsilon_{\rm rec}(B^0\rightarrow K^+\pi^-) $&$/$&$\varepsilon_{\rm rec}(B_s^0\rightarrow\pi^+ K^-)$ & $0.98 \pm 0.02$ \\
     \hline
     \multirow{2}{*}{C} &$\varepsilon_{\rm rec}(B^0\rightarrow \pi^+\pi^-) $&$/$&$\varepsilon_{\rm rec}(B_s^0\rightarrow\pi^+\pi^-)$ \bigstrut & $1.10 \pm 0.03$ \\
     & $\varepsilon_{\rm rec}(B_s^0\rightarrow K^+K^-) $&$/$&$\varepsilon_{\rm rec}(B^0\rightarrow K^+ K^-)$  & $0.92 \pm 0.02$ \\
   \end{tabular}
 \end{center}
\end{table}

\section{Particle identification}

In order to disentangle the various $H_b \rightarrow h^+h^{\prime-}$  decay modes, the selected $b$-hadron candidates are divided into different final states using the PID capabilities of the two RICH detectors. Different sets of PID criteria are applied to the candidates passing the three selections, with PID discrimination power increasing from selection A to selection C.
These criteria identify mutually exclusive sets of candidates. As discriminators we employ the quantities $\Delta \ln \mathcal{L}_{K\pi}$ and $\Delta \ln \mathcal{L}_{p\pi}$, or their difference $\Delta \ln \mathcal{L}_{Kp}$ when appropriate, where $\Delta \ln \mathcal{L}_{\alpha \beta}$ is the difference between the natural logarithms of the likelihoods for a given daughter particle under mass hypotheses $\alpha$ and $\beta$, respectively. In order to determine the corresponding PID efficiency for each two-body final state, a data-driven method is employed that uses $D^{*+} \rightarrow D^0(K^-\pi^+) \pi^+$ and $\Lambda \rightarrow p \pi^-$ decays as control samples.
In this analysis about 6.7 million $D^{*+}$ decays and 4.2 million $\Lambda$ decays are used.

The production and decay kinematics of the $D^0 \rightarrow K^-\pi^+$ and $\Lambda \rightarrow p \pi^-$ channels differ from those of the $b$-hadron decays under study. Since the RICH PID information is momentum dependent, a calibration procedure is performed by reweighting the $\Delta \ln \mathcal{L}_{\alpha \beta}$ distributions of true pions, kaons and protons obtained from the calibration samples, with the momentum distributions of daughter particles resulting from $H_b \rightarrow h^+h^{\prime-}$ decays. The $\Delta \ln \mathcal{L}_{\alpha \beta}$ and momentum distributions of the calibration samples and the momentum distributions of $H_b$ daughter particles are determined from data. In order to obtain background-subtracted distributions, extensive use of the \emph{sPlot} technique~\cite{Pivk:2004ty} is made. This technique requires that extended maximum likelihood fits are performed, where signal and background components are modelled. It is achieved by fitting suitable models to the distribution of the variable $\delta m=m_{K\pi\pi}-m_{K\pi}$ for $D^{*+}\rightarrow D^0(K^-\pi^+)\pi^+$ decays, to the $p \pi^-$ mass for $\Lambda \rightarrow p \pi^-$ decays and, for each of the three selections, to the invariant mass assuming the $\pi^+\pi^-$ hypothesis for $H_b \rightarrow h^+h^{\prime-}$ decays. The variables $m_{K\pi\pi}$ and $m_{K\pi}$ are the reconstructed $D^{*+}$ and $D^0$ candidate masses, respectively.
%Using the results of these fits, the \emph{sPlot} technique~\cite{Pivk:2004ty} is applied to obtain the momentum distributions of the daughter particles for $D^0$, $\Lambda$ and $b$-hadron decays and the $\Delta \ln \mathcal{L}_{\alpha \beta}$ distributions for pions, kaons and protons from $D^0$ and $\Lambda$ decays.

\begin{figure}[t]
  \begin{center}
    \includegraphics[width=0.48\textwidth]{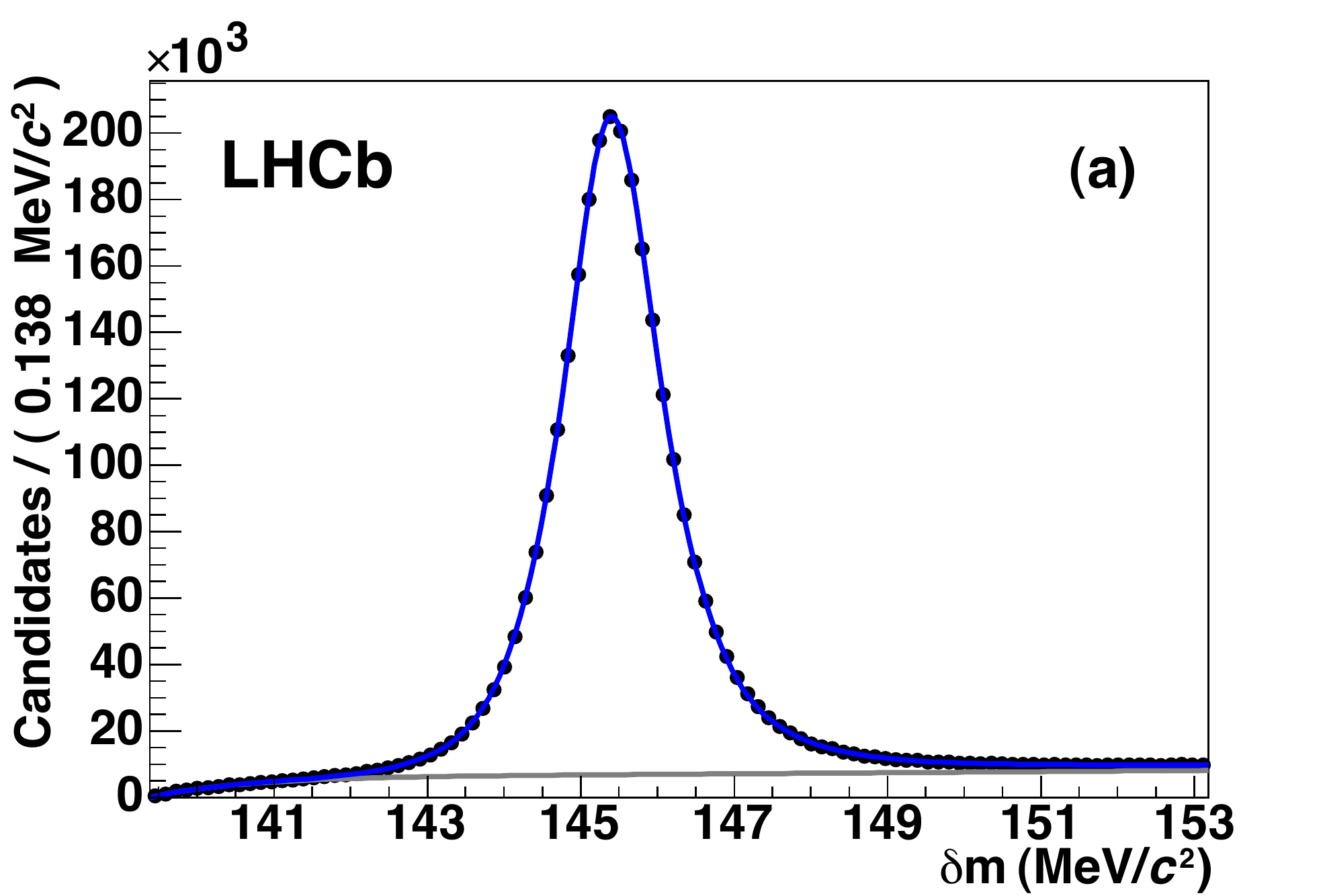}
    \includegraphics[width=0.48\textwidth]{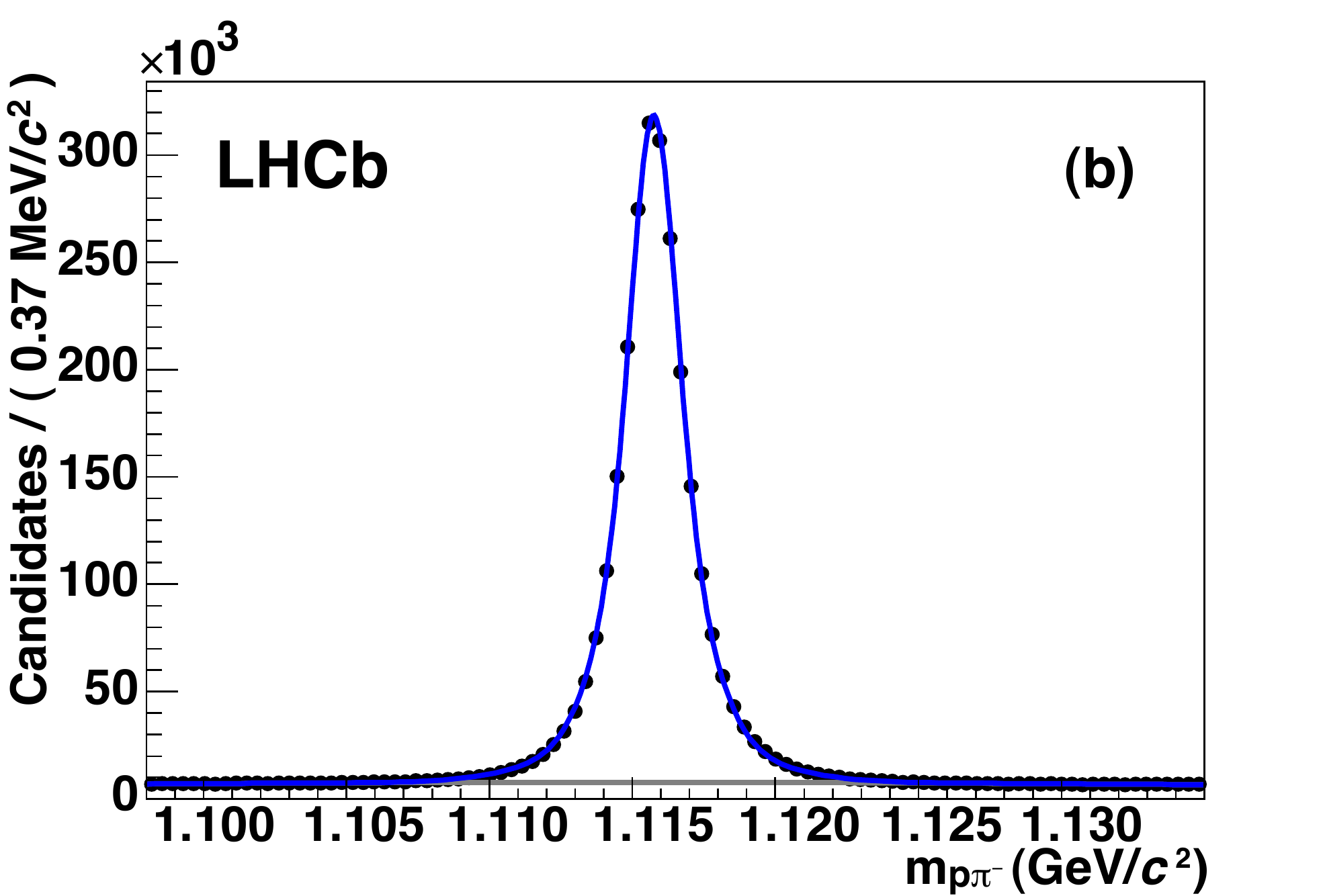}
 \end{center}
  \caption{Distributions of (a) $\delta m=m_{K\pi\pi}-m_{K\pi}$ for $D^{*+}\rightarrow D^0(K^-\pi^+)\pi^+$ candidates and (b) invariant mass of $\Lambda \rightarrow p \pi^-$ candidates, used for the PID calibration. The curves are the results of maximum likelihood fits.
}
  \label{fig:d0calib}
\end{figure}

In Fig. \ref{fig:d0calib} the distributions of the variable $\delta m$ and of the invariant mass of $\Lambda \rightarrow p \pi^-$ are shown. The superimposed curves are the results of the maximum likelihood fits to the spectra.
\begin{figure}[t]
  \begin{center}
    \includegraphics[width=0.6\textwidth]{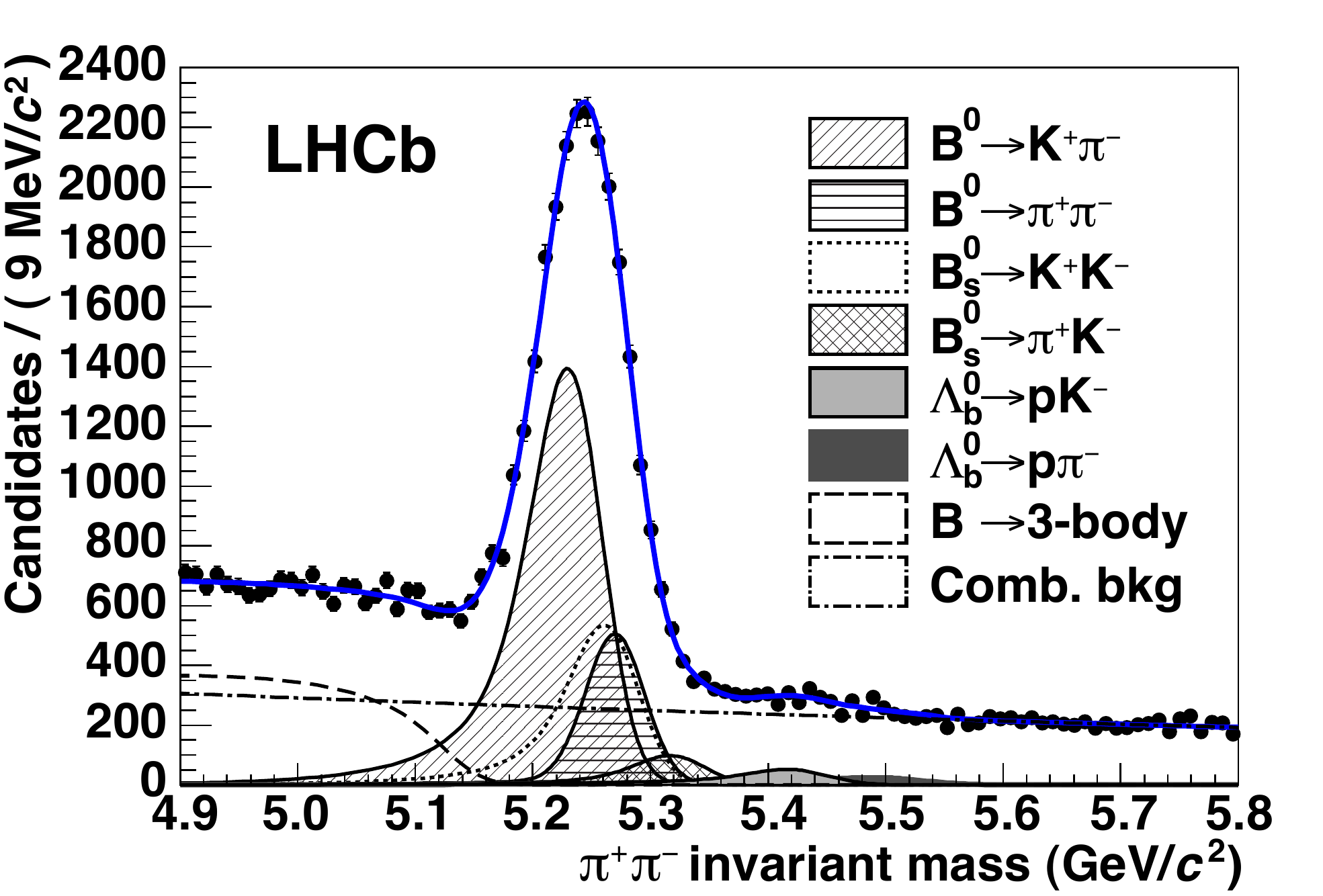}
 \end{center}
  \caption{Invariant $\pi^+\pi^-$ mass for candidates passing the selection A of Table~\ref{tab:selectioncuts}. The result of an unbinned maximum likelihood fit is overlaid. The main contributions to the fit model are also shown.}
  \label{fig:b2hhcomparisonpipi}
\end{figure}
\begin{figure}[t]
  \begin{center}
    \includegraphics[width=0.48\textwidth]{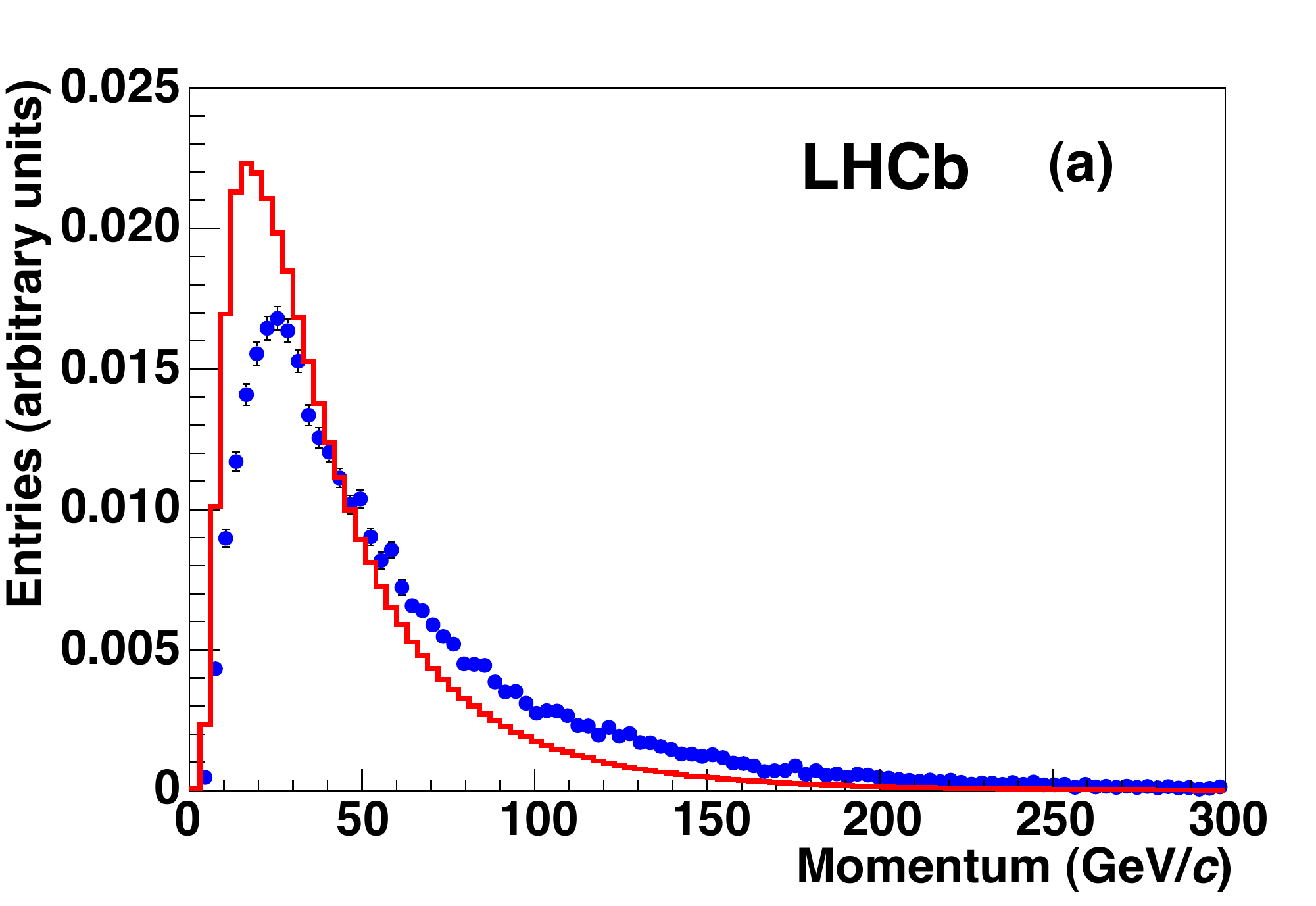}\includegraphics[width=0.48\textwidth]{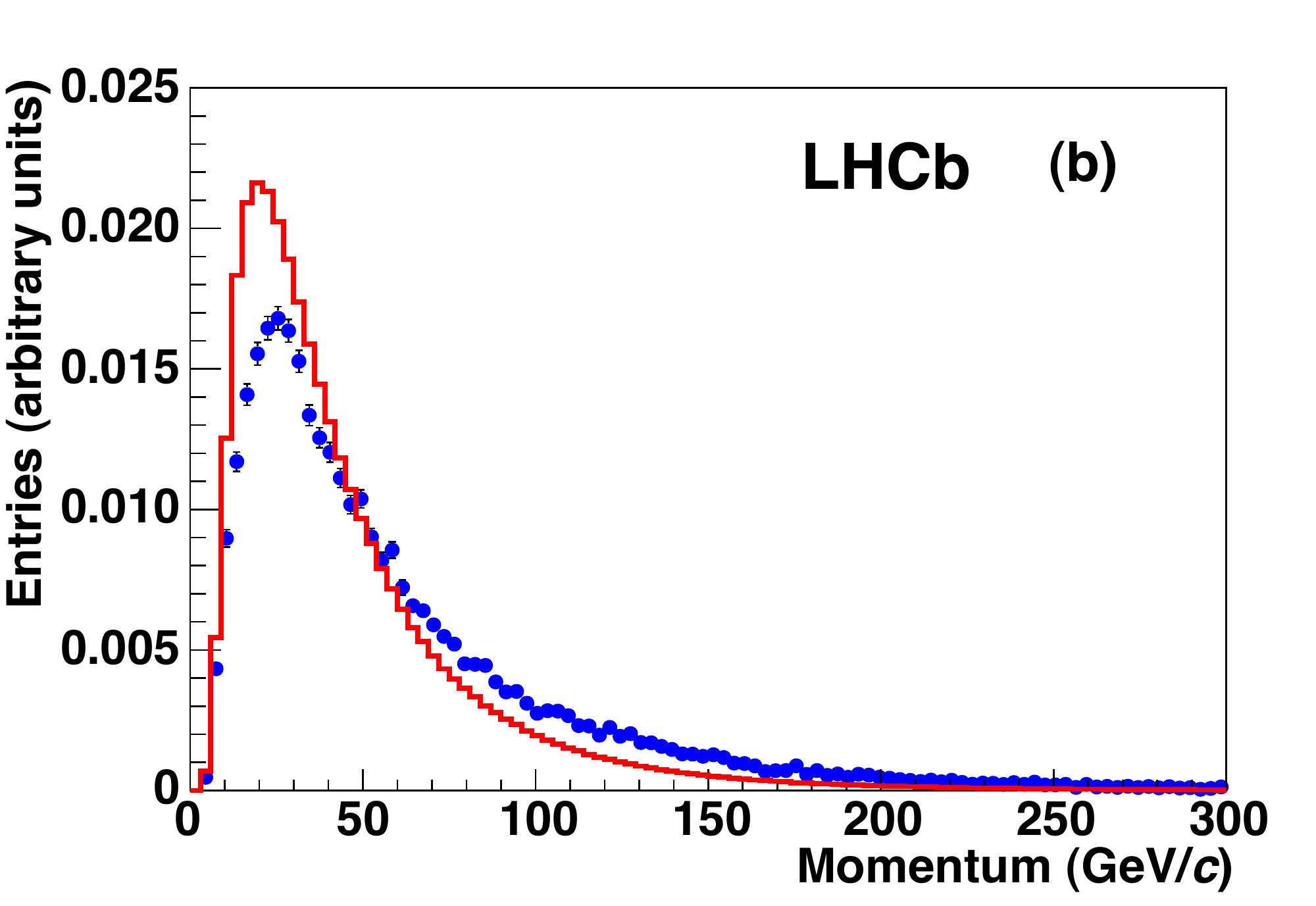}
 \end{center}
  \caption{Momentum distributions of (a) pions and (b) kaons from $D^0$ decays in the PID calibration sample (histograms). For comparison, the points represent the inclusive momentum distribution of daughter particles in $H_b \rightarrow h^+h^{\prime-}$ decays. The distributions are normalized to the same area. This example corresponds to selection A.}
  \label{fig:b2hhcomparisonmomenta}
\end{figure}
The $D^{*+}\rightarrow D^0(K^-\pi^+)\pi^+$ signal $\delta m$ spectrum has been modelled using the sum of three Gaussian functions ($G_3$) with a common mean ($\mu$), convolved with an empirical function which describes the asymmetric tail on the right-hand side of the spectrum:
\begin{equation}
  g(\delta m)=A \left [\Theta(\delta m^\prime - \mu) \cdot \left
      (\delta m^\prime - \mu\right )^s \right] 
  \otimes G_{3}(\delta m-\delta m^\prime),
  \label{eq:fitting:tagged:pdf3}
\end{equation}
where $A$ is a normalization factor, $\Theta$ is the Heaviside (step) function, $s$ is a free parameter determining the asymmetric shape of the distribution, $\otimes$ stands for convolution and the convolution integral runs over $\delta m^\prime$.
In order to model the background shape we use
\begin{equation}
  h(\delta m)=B\left[ 1- \exp \left(-\frac{\delta m-\delta m_0}{c} \right) \right ],
  \label{eq:fitting:tagged:bkg}
\end{equation}
where $B$ is a normalization factor, and the free parameters $\delta m_0$ and $c$ govern the shape of the distribution. The fit to the $\Lambda \rightarrow p \pi^-$ spectrum is made using a sum of three Gaussian functions for the signal and a second order polynomial for the background.

{
\begin{table}[t]
  \caption{PID efficiencies (in \%), for the various mass hypotheses, corresponding to the event samples passing the 
    selections A, B and C of Table~\ref{tab:selectioncuts}. Different sets of PID requirements are applied in the three cases.}
  \begin{center}
    \begin{tabular}{r@{\hspace{0.2cm}}c@{\hspace{0.2cm}}l|c|c|c|c|c}
     \multicolumn{3}{c|}{Selection A} & $\pi^+\pi^-$ & $K^+K^-$ & $K^+\pi^-$ & $p\pi^-$ & $pK^-$ \\
      \hline
      $B^0$ &$\rightarrow$ & $\pi^+\pi^-$                          &  43.1  & 0.33  & 28.6  & 1.53  & 0.13 \\
      $B^0_s$& $\rightarrow$ & $K^+K^-$                            &  0.05  & 55.0  & 15.4  & 0.05  & 1.63 \\
      $B^0_{(s)}$ &$\rightarrow$ & $K^+\pi^-$                       & 1.40  & 4.17  & 67.9  & 0.72  & 0.06 \\
      $\bar{B}^0_{(s)}$ & $\rightarrow$ & $\pi^+ K^-$              & 1.40  & 4.17  & 2.09  & 0.02  & 0.85 \\
      $\Lambda^0_b $ & $\rightarrow$ & $p\pi^-$                    & 1.93  & 0.92  & 16.8  & 35.4  & 3.16 \\
      $\bar{\Lambda}^0_b $ & $\rightarrow$ & $\pi^+ \bar{p}$  & 1.93  & 0.92  & 0.95  & 0.03  & 0.18 \\
      $\Lambda^0_b $ & $\rightarrow$ & $pK^-$                       & 0.06  & 12.2  & 1.92  & 1.18  & 40.2 \\
      $\bar{\Lambda}^0_b $ & $\rightarrow$ &$K^+ \bar{p}$ & 0.06  & 12.2  & 4.51  & 0.03  & 0.18 \\
\multicolumn{8}{c}{}\\
     \multicolumn{3}{c|}{Selection B} & $\pi^+\pi^-$ & $K^+K^-$ & $K^+\pi^-$ & $p\pi^-$ & $pK^-$ \\
      \hline
    $B^0$ &$\rightarrow$ &$\pi^+\pi^-$                          &  42.8  & 0.33 & 2.06 & 1.51 & 0.13  \\
     $B^0_s$ &$\rightarrow$ &$ K^+K^-$                            &  0.05  & 54.5 & 1.09 & 0.05 & 1.63  \\
     $B^0_{(s)}$ &$\rightarrow$ &$ K^+\pi^-$                       & 1.38  & 4.12 & 35.7 & 0.72 & 0.06 \\
     $\bar{B}^0_{(s)}$ &$\rightarrow$ &$\pi^+ K^-$              & 1.38  & 4.12 & 0.02 & 0.02 & 0.84 \\
     $\Lambda^0_b $ &$\rightarrow$ &$ p\pi^-$                    & 1.90  & 0.90 & 6.01 & 35.4 & 3.16  \\
     $\bar{\Lambda}^0_b $ &$\rightarrow$ &$\pi^+ \bar{p}$  & 1.90  & 0.90 & 0.03 & 0.03 & 0.17 \\
     $\Lambda^0_b $ &$\rightarrow$ &$ pK^-$                       & 0.06  & 11.8 & 0.09 & 1.19 & 40.2 \\
     $\bar{\Lambda}^0_b $ &$\rightarrow$ &$ K^+ \bar{p}$    & 0.06  & 11.8 & 0.88 & 0.03 & 0.17 \\
\multicolumn{8}{c}{}\\
     \multicolumn{3}{c|}{Selection C} & $\pi^+\pi^-$ & $K^+K^-$ & $K^+\pi^-$ & $p\pi^-$ & $pK^-$ \\
      \hline
       $B^0$ &$\rightarrow$ &$\pi^+\pi^-$                          &  40.5  & 0.00 & 1.64 & 1.51 & 0.00  \\
       $B^0_s$ &$\rightarrow$ &$ K^+K^-$                            &  0.04  & 21.4 & 0.98 & 0.04 & 1.01  \\
       $B^0_{(s)}$ &$\rightarrow$ &$ K^+\pi^-$                       & 1.27  & 0.11 & 32.4 & 0.70 & 0.00 \\
       $\bar{B}^0_{(s)}$ &$\rightarrow$ &$\pi^+ K^-$              & 1.27  & 0.11 & 0.01 & 0.02 & 0.54 \\
       $\Lambda^0_b $ &$\rightarrow$ &$ p\pi^-$                    & 1.26  & 0.00 & 3.16 & 33.5 & 0.13  \\
       $\bar{\Lambda}^0_b $ &$\rightarrow$ &$\pi^+ \bar{p}$  & 1.26  & 0.00 & 0.02 & 0.02 & 0.03 \\
       $\Lambda^0_b $ &$\rightarrow$ &$ pK^-$                       & 0.04  & 1.35 & 0.05 & 1.08 & 23.9 \\
       $\bar{\Lambda}^0_b $ &$\rightarrow$ &$ K^+ \bar{p}$    & 0.04  & 1.35 & 0.65 & 0.02 & 0.03 \\
   \end{tabular}
  \end{center}
 \label{tab:eff_B}
\end{table}
}

Figure~\ref{fig:b2hhcomparisonpipi} shows the invariant mass assuming
the $\pi^+\pi^-$ hypothesis for selected $b$-hadron candidates, using the kinematic selection A of Table~\ref{tab:selectioncuts} and without applying any PID requirement. The shapes describing the various signal decay modes have been fixed by parameterizing the mass distributions obtained from Monte
Carlo simulation convolved with a Gaussian resolution
function with variable mean and width.  The three-body and combinatorial
backgrounds are modelled using an ARGUS function~\cite{Albrecht:1989ga}, convolved with the same Gaussian resolution function used for the signal distributions, and an exponential function,
respectively.
The relative yields between the signal components have
been fixed according to the known values of branching fractions and
hadronization probabilities of $B^0$, $B^0_s$ and $\Lambda^0_b$ hadrons~\cite{bib:hfagbase}.
The fits corresponding to the kinematic selection criteria B and C of Table~\ref{tab:selectioncuts} have also been made, although not shown, in order to take into account possible differences in the momentum distributions due to different selection criteria.

{
\renewcommand{\arraystretch}{1.1}
\begin{table}[t]
    \caption{Ratios of PID efficiencies used to compute the relevant ratios of branching fractions, corresponding to selection A.}
  \begin{center}
    \begin{tabular}{r@{\hspace{0.1cm}}c@{\hspace{0.1cm}}l|c}
     \multicolumn{3}{c|}{Efficiency ratio} & Value \\
      \hline
      $\varepsilon_{\rm PID}(K^+\pi^-)$&$/$&$\varepsilon_{\rm PID}(\pi^+\pi^-)$ & $1.57 \pm 0.09$\\
     $\varepsilon_{\rm PID}(K^+\pi^-) $&$/$&$\varepsilon_{\rm PID}(K^+K^-)$ & $1.23 \pm 0.06$ \\
     $\varepsilon_{\rm PID}(p K^-) $&$/$&$\varepsilon_{\rm PID}(p \pi^-)$ & $1.14 \pm 0.05$ \\
   \end{tabular}

\label{tab:pidratiobd}
  \end{center}
\end{table}
}

As mentioned above, the \emph{sPlot} procedure is used to determine the various $\Delta \ln \mathcal{L}_{\alpha \beta}$ and momentum distributions, and these are used to reweight the $D^{*+}$ and $\Lambda$ calibration samples.
As an example, the momentum distributions of pions and kaons from $D^0$ decays and the inclusive momentum distribution of daughter particles in $H_b \rightarrow h^+h^{\prime-}$ decays, the latter corresponding to selection A, are shown in Fig.~\ref{fig:b2hhcomparisonmomenta}.

The PID efficiencies corresponding to the three selections are determined by applying the PID selection criteria to the reweighted $D^{*+}$ and $\Lambda$ calibration samples. The results are reported in Table~\ref{tab:eff_B}.
Using these efficiencies, the relevant PID efficiency ratios are determined and summarized in Table~\ref{tab:pidratiobd}. These ratios correspond to selection A only, since for the measurements involved in B and C the final states are identical and the ratios of PID efficiencies are equal to unity. It has been verified that the PID efficiencies do not show any sizeable dependence on the flavour of the parent hadron, as differences in the momentum distributions of the daughter particles for different parent hadrons are found to be small.
Owing to the large sizes of the calibration samples, the uncertainties associated to the PID efficiency ratios are dominated by systematic effects, intrinsically related to the calibration procedure. They are estimated by means of a data-driven approach, where several fits to the $B^0 \to K^+\pi^-$ mass spectrum are made. The mass distributions in each fit are obtained by varying the PID selection criteria over a wide range, and then comparing the variation of the $B^0 \to K^+\pi^-$ signal yields determined by the fits to that of the PID efficiencies predicted by the calibration procedure. The largest deviation is then used to estimate the size of the systematic uncertainty.

\section{Invariant mass fits to {\boldmath $H_b \rightarrow h^+h^{\prime -}$} spectra}
\label{sec:fittingModel}

\begin{figure}[t]
\begin{center}
\includegraphics[width=0.48\textwidth]{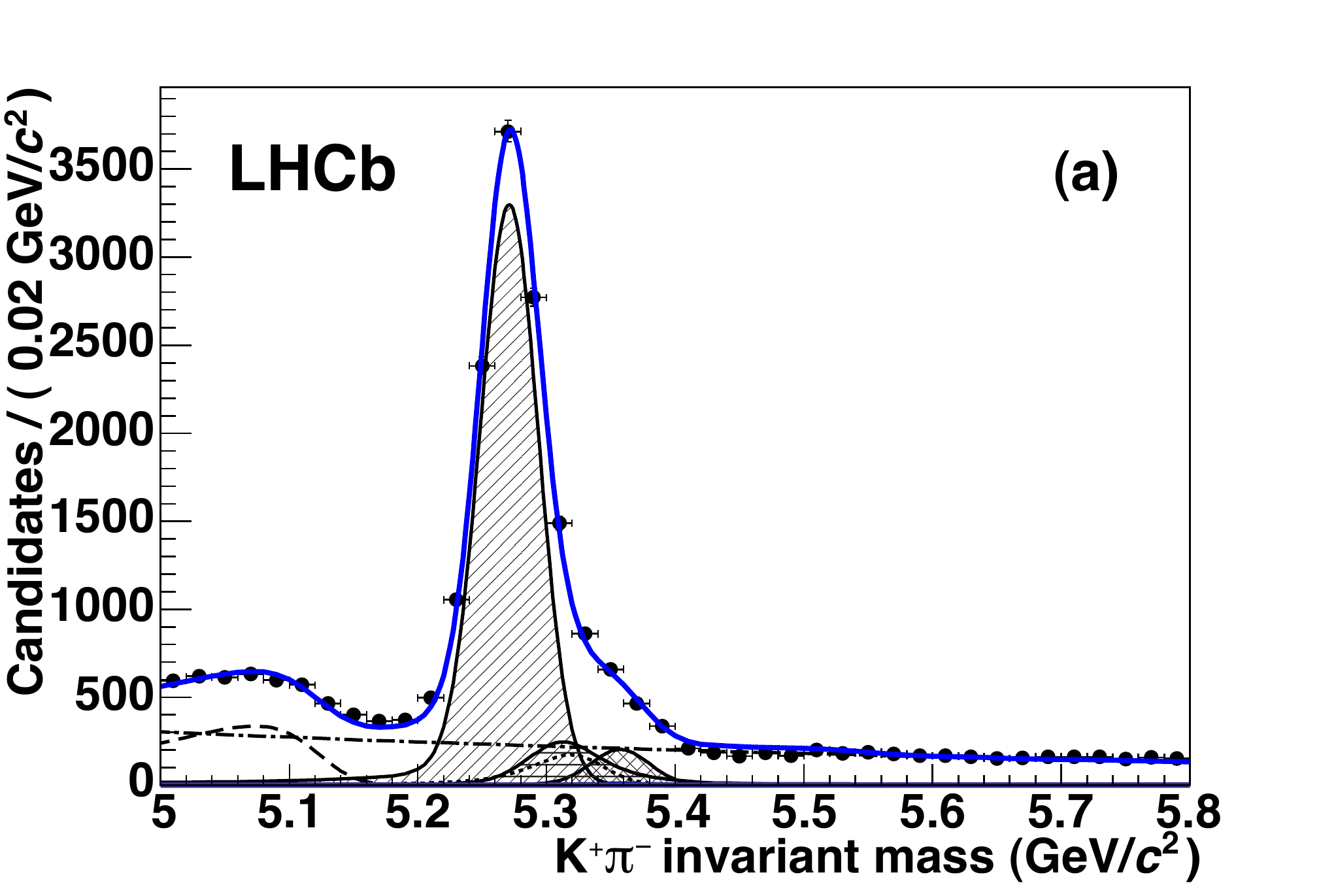}
\includegraphics[width=0.48\textwidth]{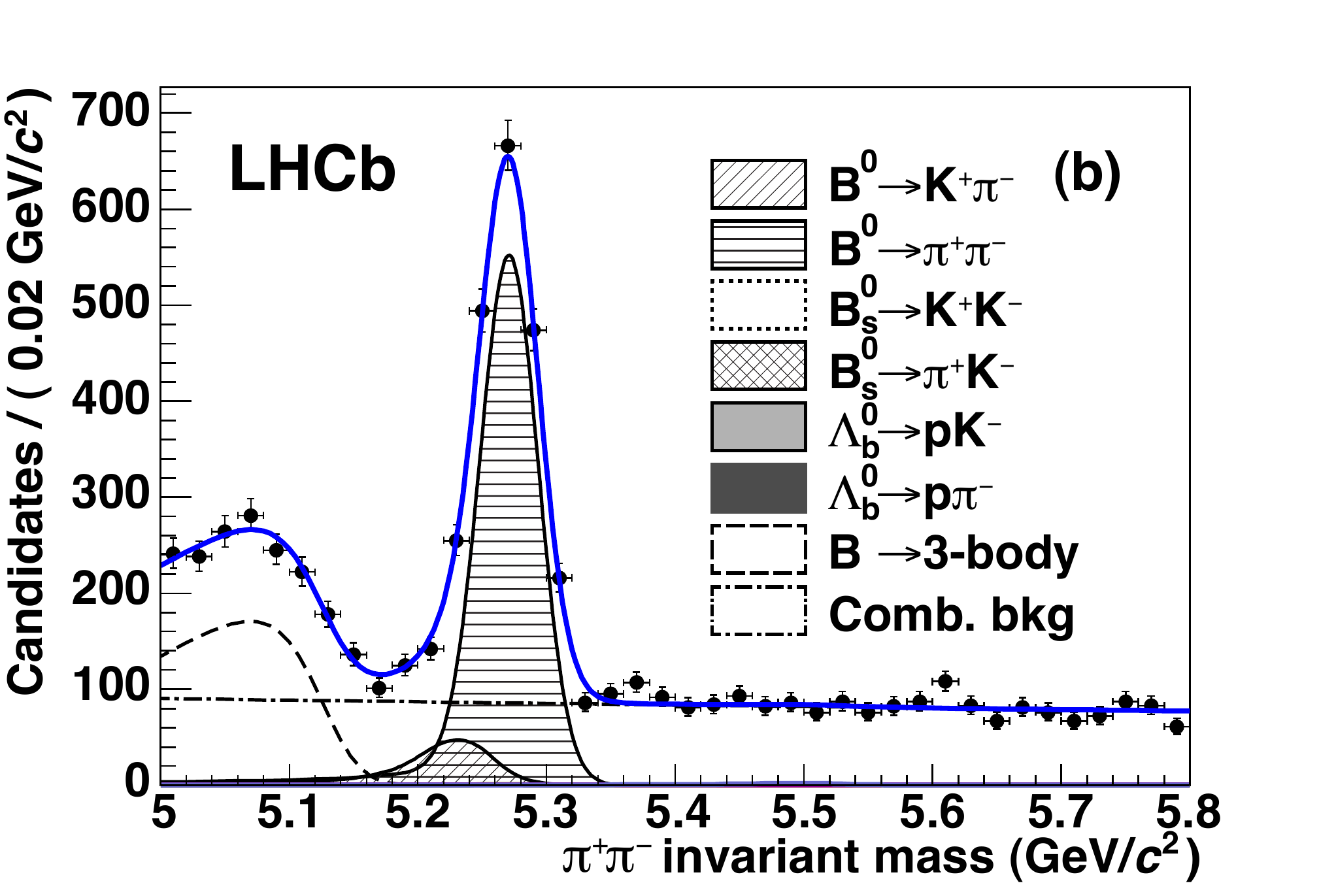}
\includegraphics[width=0.48\textwidth]{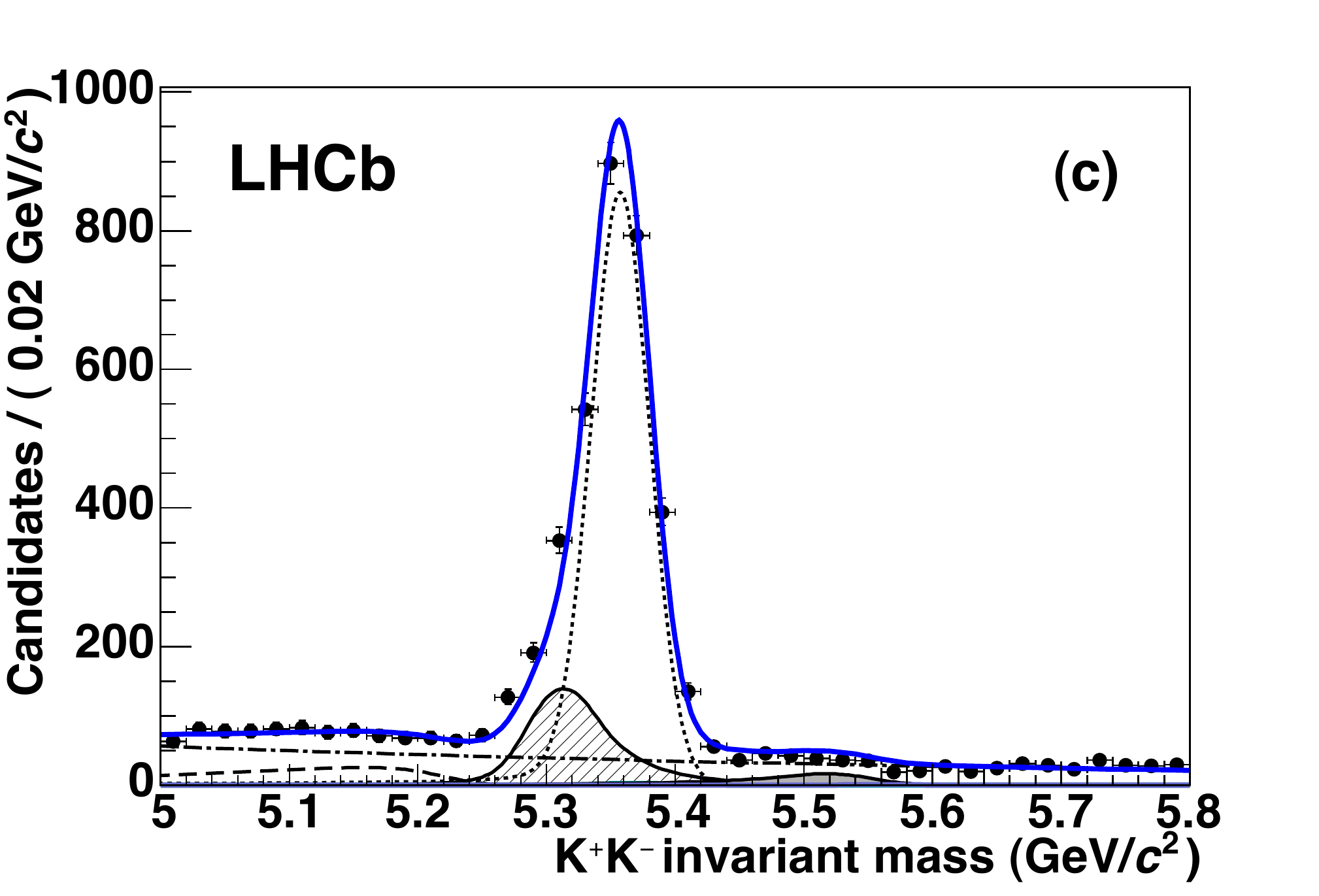}
\includegraphics[width=0.48\textwidth]{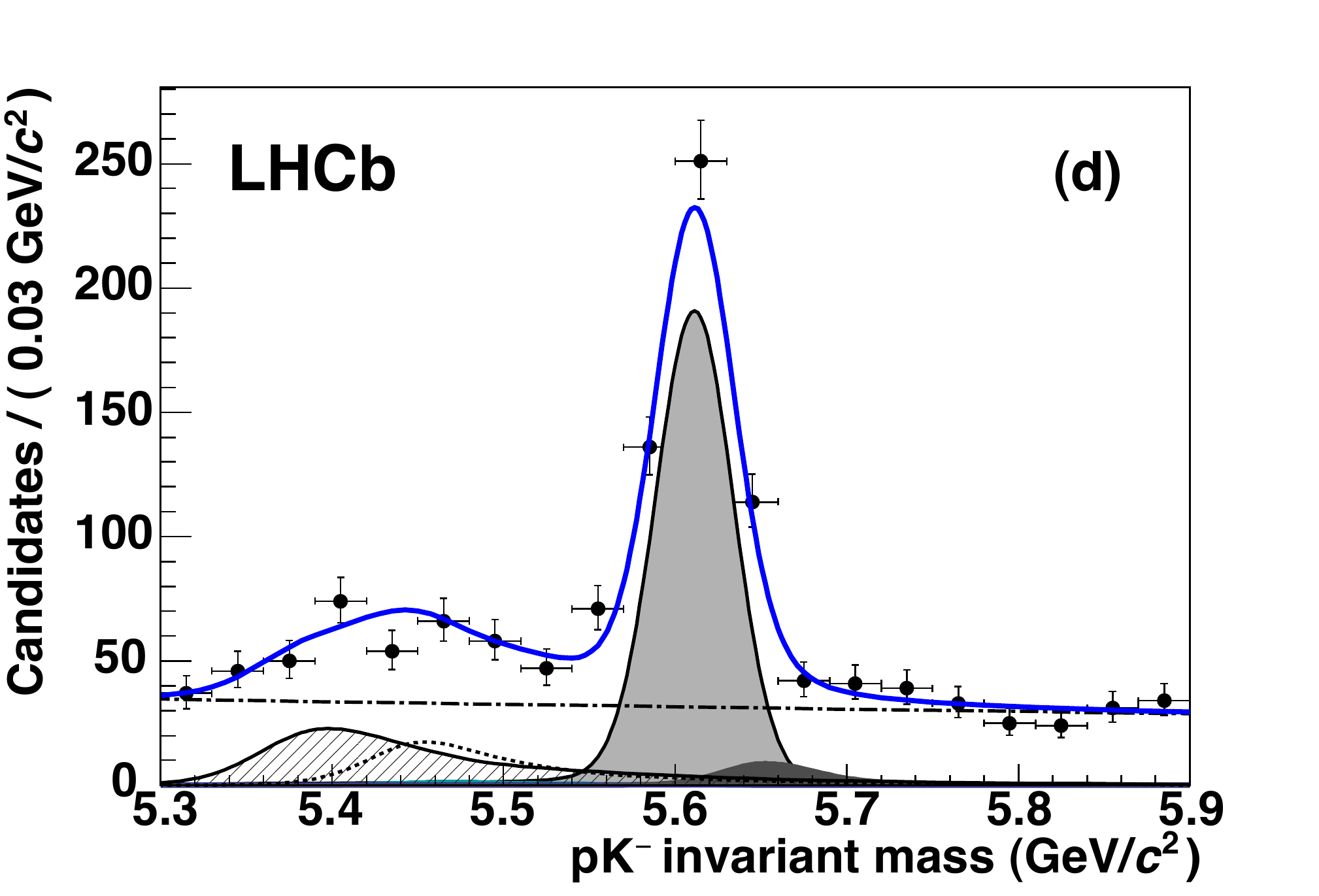}
\includegraphics[width=0.48\textwidth]{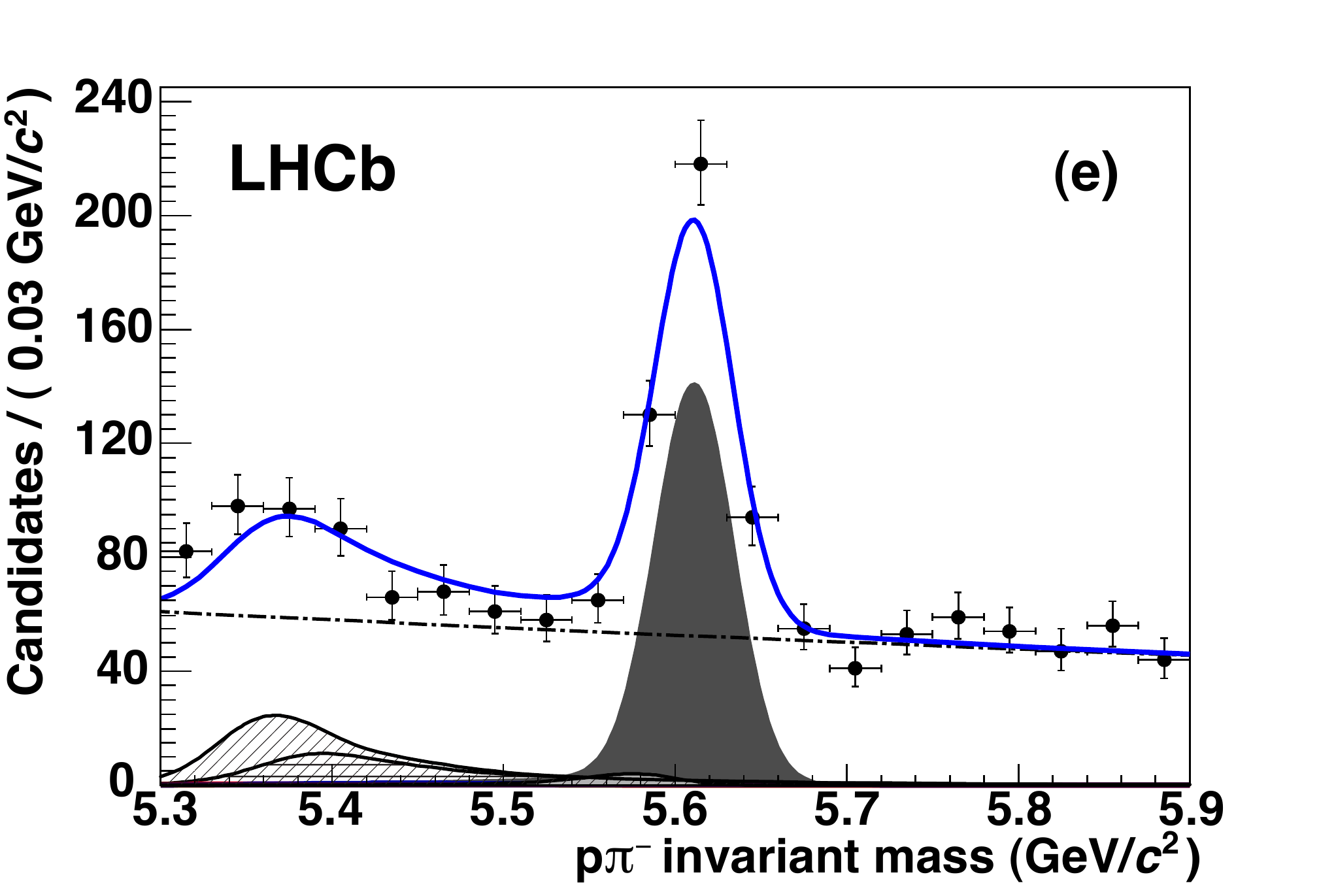}
\includegraphics[width=0.48\textwidth]{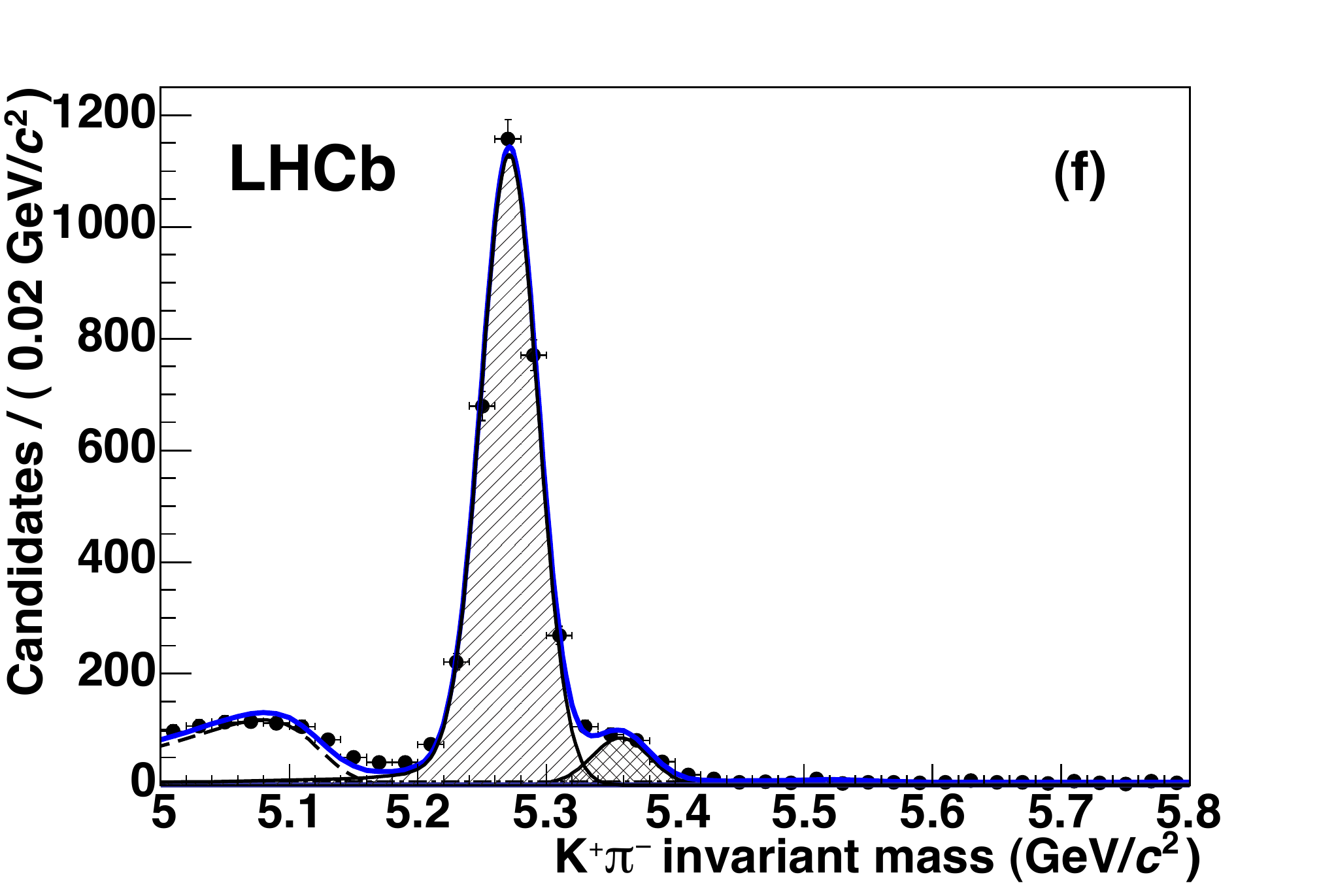}
\end{center}
\caption{Invariant mass spectra corresponding to selection A for the mass hypotheses (a) $K^+\pi^-$, (b) $\pi^+\pi^-$, (c) $K^+K^-$, (d) $pK^-$ and (e) $p\pi^-$, and to selection B for the mass hypothesis (f) $K^+\pi^-$. The results of the unbinned maximum likelihood fits are overlaid. The main components contributing to the fit model are also shown.}
\label{fig:bdbsfit}
\end{figure}

Unbinned maximum likelihood fits are performed to the mass spectra of events passing the selections A, B and C with associated PID selection criteria. For each selection we have five different spectra, corresponding to the final state hypotheses $K^+\pi^-$, $\pi^+\pi^-$, $K^+K^-$, $pK^-$ and $p\pi^-$, to which we perform a simultaneous fit. Since each signal channel is also a background for all the other signal decay modes in case of misidentification of the final state particles (cross-feed background), the simultaneous fits to all the spectra allow a determination of the yields of the signal components together with those of the cross-feed backgrounds, once the appropriate PID efficiency factors are taken into account. The signal component for each hypothesis is described by a single Gaussian distribution, convolved with a function which describes the effect of the final state radiation on the mass line shape~\cite{Baracchini:2005wp}. The combinatorial background is modelled by an exponential function and the shapes of the cross-feed backgrounds are obtained from Monte Carlo simulation. The background due to partially reconstructed three-body $B$ decays is parameterized by an ARGUS function~\cite{Albrecht:1989ga} convolved with a Gaussian resolution function that has the same width as the signal distribution. 

\begin{figure}[t]
\begin{center}
\includegraphics[width=0.48\textwidth]{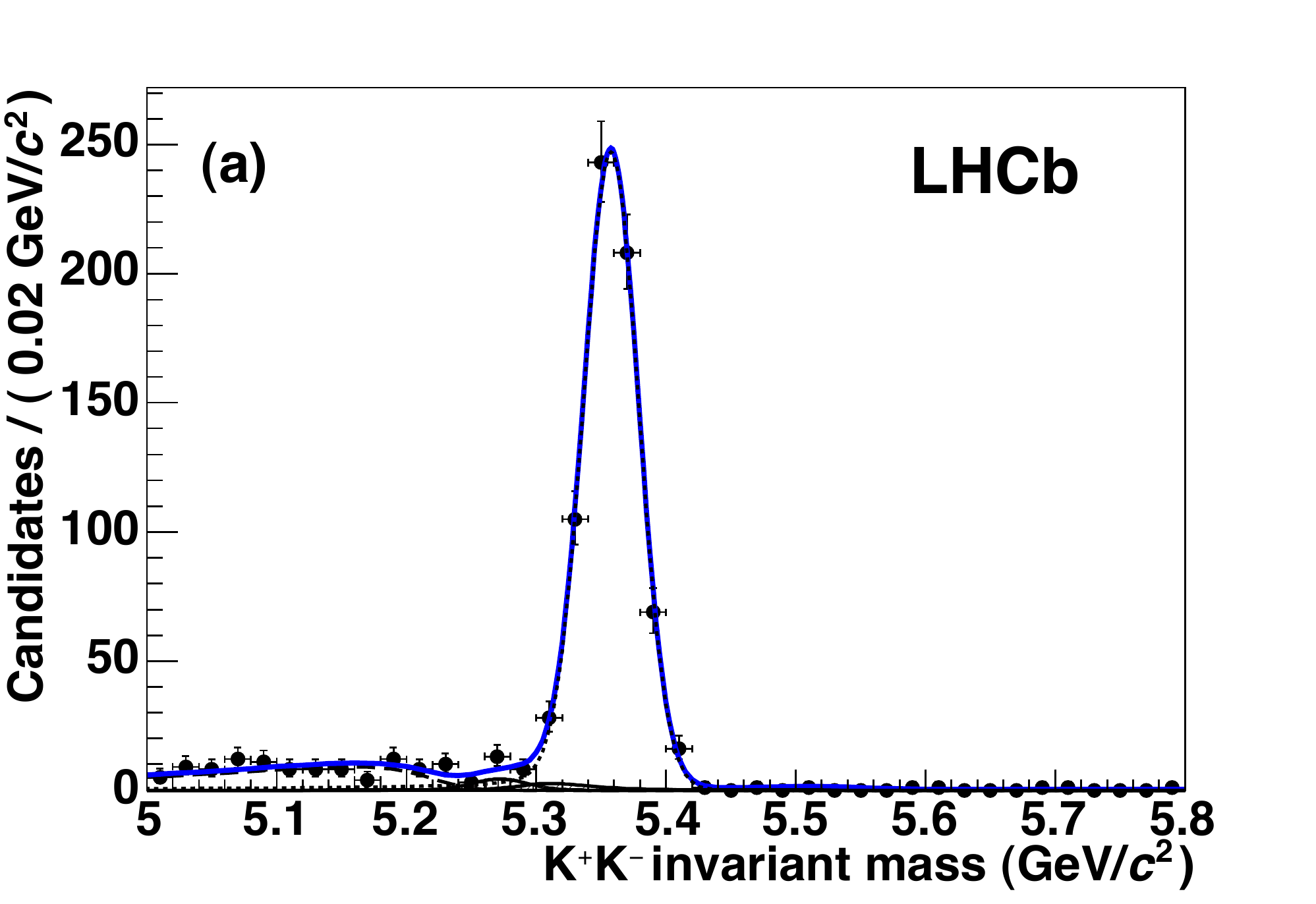} \includegraphics[width=0.48\textwidth]{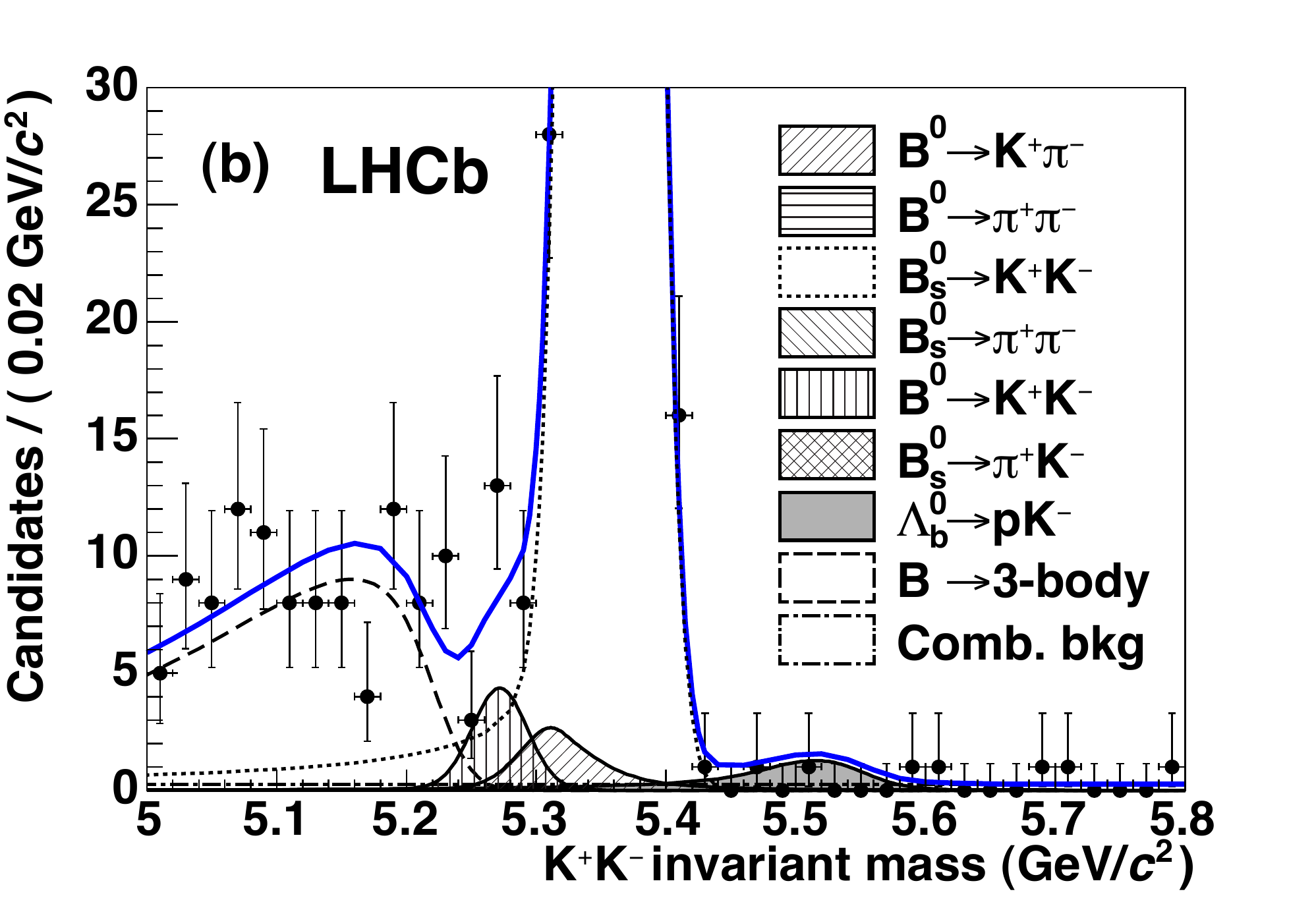}
\includegraphics[width=0.48\textwidth]{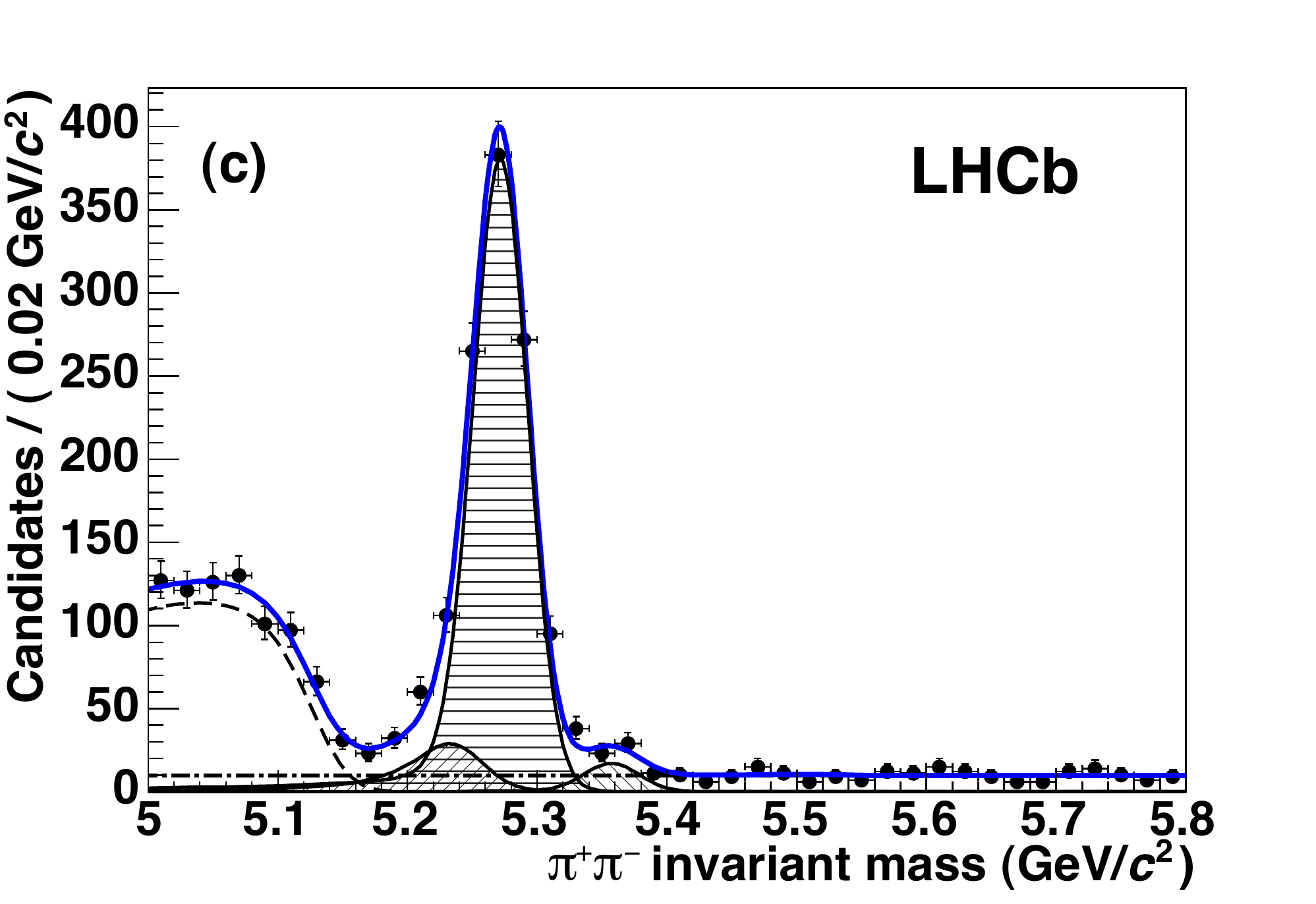} \includegraphics[width=0.48\textwidth]{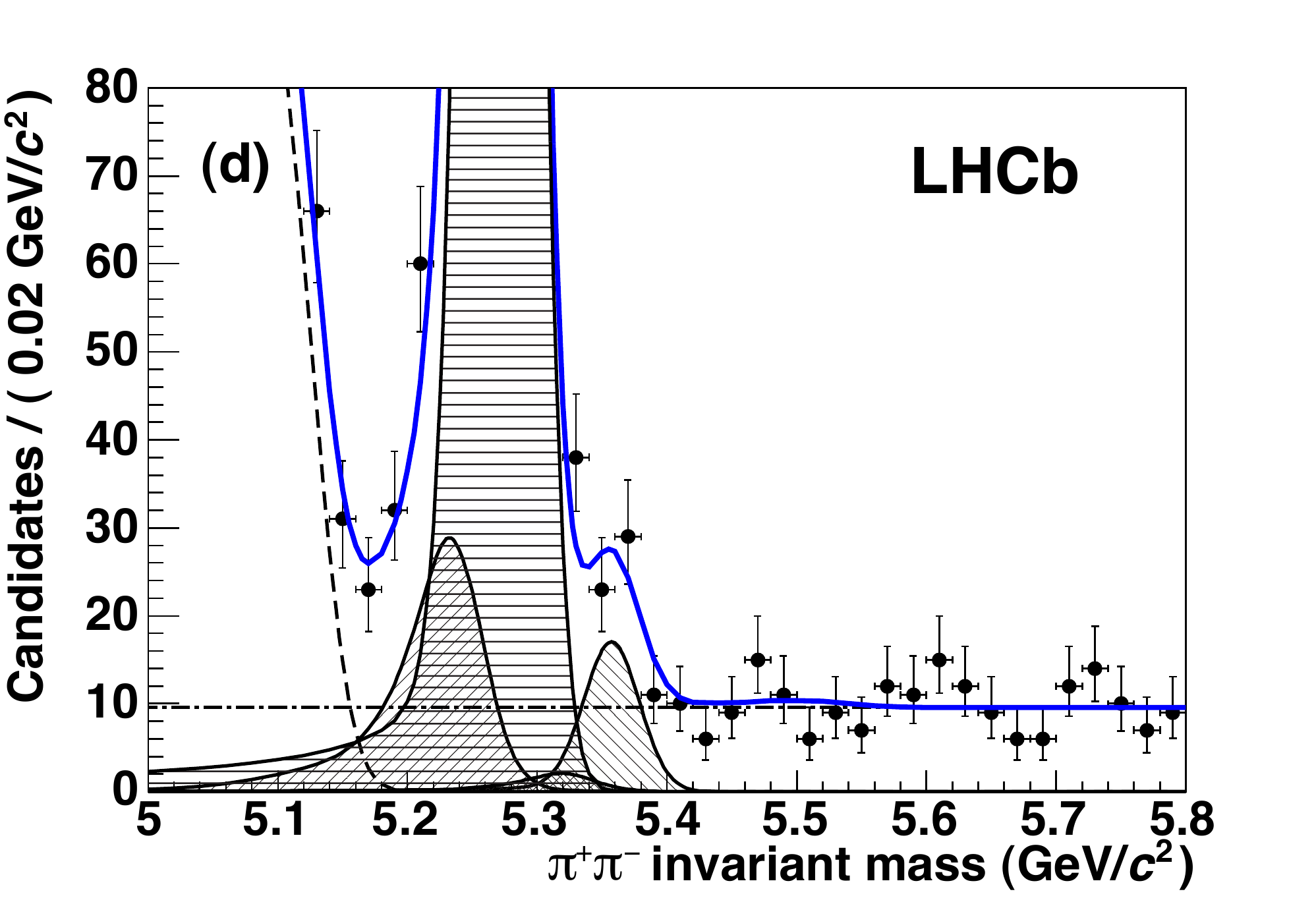}
\end{center}
\caption{Invariant mass spectra corresponding to selection C for the mass hypotheses (a, b) $K^+K^-$ and (c, d) $\pi^+\pi^-$. Plots (b) and (d) are the same as (a) and (c) respectively, but magnified to focus on the rare $B^0 \rightarrow K^+ K^-$ and $B^0_s \rightarrow \pi^+ \pi^-$ signals. The results of the unbinned maximum likelihood fits are overlaid. The main components contributing to the fit model are also shown.}
\label{fig:raredecays}
\end{figure}

{
\renewcommand{\arraystretch}{1.1}
\begin{table}[t]
 \caption{Signal yields determined by the unbinned maximum likelihood fits to the data samples surviving the event selections A, B and C of Table~\ref{tab:selectioncuts} with the associated PID criteria. Only statistical uncertainties are shown.}
  \begin{center}
    \begin{tabular}{c|r@{\hspace{0.2cm}}c@{\hspace{0.2cm}}l|r@{\hspace{0.1cm}}c@{\hspace{0.1cm}}l}
     Selection & \multicolumn{3}{c|}{Decay} & \multicolumn{3}{c}{Signal yield} \tabularnewline
      \hline 
      \multirow{5}{*}{A} & $B^{0}$&$ \rightarrow $&$ K^+\pi^-$ & $ 9822 $&$ \pm $&$ 122 $ \tabularnewline
                                             & $B^{0}$&$ \rightarrow $&$ \pi^+\pi^-$ & $ 1667 $&$ \pm $&$ 51 $ \tabularnewline
                                              & $B^{0}_s $&$ \rightarrow $&$ K^+K^-$  & $ 2523 $&$ \pm $&$ 59 $ \tabularnewline
                                              & $\Lambda^0_b $&$ \rightarrow $&$ pK^-$  & $ 372 $&$ \pm $&$ 22 $ \tabularnewline
                                             & $\Lambda^0_b $&$ \rightarrow $&$ p\pi^-$  & $ 279 $&$ \pm $&$ 22 $ \tabularnewline
      \hline 
     \multirow{2}{*}{B} & $B^{0}$&$ \rightarrow $&$ K^+\pi^-$ & $ 3295 $&$ \pm $&$ 59 $ \tabularnewline
                                  & $B^{0}_s $&$ \rightarrow $&$ \pi^+ K^-$ & $ 249 $&$ \pm $&$ 20 $ \tabularnewline
      \hline
       \multirow{4}{*}{C} & $B^{0}$&$ \rightarrow $&$ \pi^+\pi^-$ & $1076 $&$ \pm $&$ 36$ \tabularnewline
                                    & $B^{0}_s $&$ \rightarrow $&$ K^+K^-$  & $682 $&$ \pm $&$ 27$ \tabularnewline
                                   & $B^{0}$&$ \rightarrow $&$ K^+K^-$ & \multicolumn{3}{c}{$13\,^{+\,6}_{-\,5}$} \tabularnewline
                                   &$B^{0}_s $&$ \rightarrow $&$ \pi^+\pi^-$  & \multicolumn{3}{c}{$ 49 \,^{+\,11}_{-\,9} $} \tabularnewline
   \end{tabular}
    \label{tab:fit_results}
  \end{center}
\end{table}
}
{
\renewcommand{\arraystretch}{1.3}
\begin{table}[t]
    \caption{Ratios of signal yields needed for the measurement of the relative branching fractions. Only statistical uncertainties are shown.}
    \label{tab:relativeYield}
  \begin{center}
    \begin{tabular}{c|c|c}
     Selection & Ratio & Value \\
      \hline
\bigstrut      \multirow{3}{*}{A} & $\frac{N(B^0\rightarrow\pi^+\pi^-)}{N(B^0\rightarrow K^+\pi^-)}$ & $0.170 \pm 0.006$ \\[1ex]
                                   & $\frac{N(B_s^0\rightarrow K^+ K^-)}{N(B^0\rightarrow K^+\pi^-)}$ & $0.257 \pm 0.007$ \\[1ex]
                                  & $\frac{N(\Lambda^0_b\rightarrow p\pi^-)}{N(\Lambda^0_b\rightarrow p K^-)}$ & $0.75 \pm 0.07$ \\[1ex]
      \hline
\bigstrut     \multirow{1}{*}{B} & $\frac{N(B_s^0\rightarrow\pi^+ K^-)}{N(B^0\rightarrow K^+\pi^-)} $ & $0.076 \pm 0.006 $ \\[1ex]
     \hline
\bigstrut      \multirow{2}{*}{C} & $\frac{N(B^0\rightarrow K^+K^-)}{N(B^0_s\rightarrow K^+K^-)}$ & $0.019 \,^{+\,0.009}_{-\,0.007}$ \\[1ex]
                                   & $\frac{N(B_s^0\rightarrow \pi^+ \pi^-)}{N(B^0\rightarrow \pi^+\pi^-)}$ & $0.046 \,^{+\,0.010}_{-\,0.009}$ \\[1ex]
   \end{tabular}
  \end{center}
\end{table}
}

The overall mass resolution determined from the fits is about 22~\mevcc. Figure~\ref{fig:bdbsfit} shows the $K^+\pi^-$, $\pi^+\pi^-$, $K^+K^-$, $pK^-$ and $p\pi^-$ invariant mass spectra corresponding to selection A and the $K^+\pi^-$ spectrum corresponding to selection B. 
Figure \ref{fig:raredecays} shows the $\pi^+\pi^-$ and $K^+K^-$ mass spectra corresponding to selection C. As is apparent in the latter, while a $B^0_s \rightarrow \pi^+\pi^-$ mass peak is visible above the combinatorial background, there are not yet sufficient data to observe a significant $B^0 \rightarrow K^+K^-$ signal. As an additional complication, the mass peak of the $B^0 \rightarrow K^+K^-$ decay is expected in a region where various components give non-negligible contributions, in particular the radiative tail of the $B^0_s \rightarrow K^+K^-$ decay and the $B^0 \rightarrow K^+\pi^-$ cross-feed background.
The relevant event yields for each of the three selections are summarized in Table \ref{tab:fit_results}. Using the values listed in Table~\ref{tab:fit_results}, we can calculate the ratios of yields needed to compute the relative branching fractions. These ratios are given in Table~\ref{tab:relativeYield}, with their statistical uncertainties.

\section{Systematic uncertainties}

{
\renewcommand{\arraystretch}{1.1}
\begin{table}[t]
   \caption{Systematic uncertainties on the ratios of signal yields. The total systematic uncertainties are obtained by summing the individual contributions in quadrature.}\label{tab:systFit1}
\vspace{-0.5cm}
  \begin{center}
\resizebox{\textwidth}{!}{
    \begin{tabular}{c|c|c|c|c|c|c}
       Syst. uncertainty  &   $\frac{N(B^0 \rightarrow \pi^+\pi^-)}{N(B^0\rightarrow K^+\pi^-)}$ \bigstrut &  $\frac{N(B^0_s \rightarrow K^+ K^-)}{N(B^0\rightarrow K^+\pi^-)}$ & $\frac{N(\Lambda^0_b\rightarrow p \pi^-)}{N(\Lambda^0_b\rightarrow p K^-)}$ & $\frac{N(B_s^0\rightarrow \pi^+ K^-)}{N(B^0\rightarrow K^+\pi^-)}$  & $\frac{N(B^0 \rightarrow  K^{+} K^{-})}{N(B^0_s \rightarrow  K^{+} K^{-})}$ & $\frac{N(B^0_s\rightarrow \pi^{+} \pi^{-})}{N(B^0\rightarrow \pi^{+} \pi^{-})}$ \\
        \hline
        PID calibration                                                   & $0.0002$ & $0.0012$ & $0.0075$ & $0.0013$ & $0.0005$ & $0.0002$\\
       Final state rad.                                           & $0.0019$ & $0.0043$ & $0.0140$& $0.0012$ & $0.0093$   & $0.0013$\\
       Signal model                                                      & negligible & $0.0001$ & $0.0013$  & $0.0052$ & $0.0010$   & $0.0031$\\
       Comb. bkg model                     & $0.0013$ & $0.0006$& $0.0086$  & negligible & $0.0012$ & $0.0004$ \\
       $K\pi$ 3-body bkg                               & $0.0018$  & $0.0048$ & $0.0239$ & $0.0011$ & negligible & negligible \\
       Cross-feed bkg               & $0.0023$ & $0.0045$& $0.0042$  & $0.0008$ & $0.0008$ & $0.0002$ \\
       \hline
        Total                                                                 & $0.0038$  & $0.0080$ & $0.0304$ & $0.0056$ &   $0.0095$     & $0.0034$ \\
\end{tabular}
}
  \end{center}
\end{table}
}

The systematic uncertainties on the ratios of signal yields are related to the PID calibration and to the modelling of the signal and background components in the maximum likelihood fits.
Knowledge of PID efficiencies is necessary to compute the number of cross-feed background events affecting the fit of any $H_b$ mass spectrum. In order to estimate the impact of imperfect PID calibration, we perform unbinned maximum likelihood fits after having altered the number of cross-feed background events present in the relevant mass spectra according to the systematic uncertainties affecting the PID efficiencies.
An estimate of the uncertainty due to possible imperfections in the description of the final state radiation is determined by varying, over a wide range, the amount of emitted
radiation~\cite{Baracchini:2005wp} in the signal line shape parameterization.  The possibility of an incorrect description of the core distribution in the signal mass model is investigated by replacing the single Gaussian with the sum of two Gaussian functions with a common mean.  The impact of additional three-body $B$ decays in the $K^+\pi^-$ spectrum, not accounted for in the baseline fit --- namely $B\rightarrow \pi\pi\pi$ where one pion is missed in the reconstruction and another is misidentified as a kaon --- is investigated. The mass line shape of this background component is determined from Monte Carlo simulation, and the fit is repeated after having modified the baseline parameterization accordingly.
For the modelling of the combinatorial background component, the fit is repeated using a first-order polynomial. Finally, for the cross-feed backgrounds, two distinct systematic uncertainties are estimated: one due to a relative bias in the mass scale of the simulated distributions with respect to the signal distributions in data, and another accounting for the difference in mass resolution between simulation and data. All the shifts from the relevant baseline values are accounted for as systematic uncertainties.
A summary of all systematic uncertainties on the ratios of event yields is reported in Table~\ref{tab:systFit1}. The total uncertainties are obtained by summing the individual contributions in quadrature. The uncertainties on the ratios of reconstruction and PID efficiencies, reported in Tables~\ref{tab:recratio} and~\ref{tab:pidratiobd}, are also included in the computation of the total systematic uncertainties on the ratios of branching fractions, reported in the next section.

\section{Results and conclusions}
\label{sec:syst}

The following quantities are determined using Eq.\,(\ref{eq:brformula}) and the values reported in
Tables~\ref{tab:recratio},~\ref{tab:pidratiobd},~\ref{tab:relativeYield} and~\ref{tab:systFit1}:

\begin{eqnarray*}
\mathcal{B}\left(B^{0}\rightarrow\pi^{+}\pi^{-}\right) /\,\mathcal{B}\left(B^{0}\rightarrow K^+\pi^-\right) & = & 0.262\pm 0.009\pm 0.017,\\
(f_{s} / f_{d}) \cdot \mathcal{B}\left(B^{0}_{s}\rightarrow K^{+}K^{-}\right) /\, \mathcal{B}\left(B^{0}\rightarrow K^+\pi^-\right) & = & 0.316\pm 0.009\pm 0.019,\\
(f_{s} / f_{d}) \cdot \mathcal{B}\left(B^0_{s}\rightarrow\pi^+ K^-\right) /\, \mathcal{B}\left(B^{0}\rightarrow K^+\pi^-\right) & = & 0.074 \pm 0.006\pm 0.006,\\
(f_{d} / f_{s}) \cdot \mathcal{B}\left(B^{0} \rightarrow K^{+}K^{-}\right) /\, \mathcal{B}\left(B^{0}_s\rightarrow K^+K^-\right) & = & 0.018 \,^{+\,0.008}_{-\,0.007} \pm 0.009,\\
(f_{s} / f_{d}) \cdot \mathcal{B}\left(B^{0}_{s}\rightarrow \pi^{+}\pi^{-}\right) /\, \mathcal{B}\left(B^{0}\rightarrow \pi^+\pi^-\right) & = & 0.050 \,^{+\,0.011}_{-\,0.009} \pm 0.004,\\
\mathcal{B}\left(\Lambda^0_b\rightarrow p\pi^-\right) /\, \mathcal{B}\left(\Lambda^0_b\rightarrow pK^-\right) & = & 0.86 \pm 0.08\pm 0.05,
\end{eqnarray*} 
where the first uncertainties are statistical and the second systematic.
Using the current world average $\mathcal{B}(B^0\rightarrow K^+\pi^-)=(19.4 \pm 0.6)\times 10^{-6}$ provided by the Heavy Flavor Averaging Group~\cite{bib:hfagbase}, 
and our measurement of the ratio between the $b$-quark hadronization probabilities $f_{s}/f_{d}=0.267\,^{+\,0.021}_{-\,0.020}$~\cite{Aaij:2011jp},
we obtain the following branching fractions:

\begin{eqnarray*}
\mathcal{B}\left(B^{0}\rightarrow\pi^{+}\pi^{-}\right) & = & (5.08 \pm 0.17 \pm 0.37)\times 10^{-6},\\
\mathcal{B}\left(B^{0}_{s}\rightarrow K^{+}K^{-}\right) & = & (23.0 \pm 0.7 \pm  2.3)\times 10^{-6},\\
\mathcal{B}\left(B^0_{s}\rightarrow\pi^+ K^-\right) & = & (5.4 \pm 0.4 \pm  0.6)\times 10^{-6},\\
\mathcal{B}(B^0 \rightarrow K^+K^-) & = & (0.11 \,^{+\,0.05}_{-\,0.04} \pm 0.06)\times10^{-6},\\
\mathcal{B}(B^0_s \rightarrow \pi^+\pi^-) & = & (0.95 \,^{+\,0.21}_{-\,0.17} \pm 0.13)\times10^{-6},
\end{eqnarray*} 
where the systematic uncertainties include the uncertainties on $\mathcal{B}(B^0\rightarrow K^+\pi^-)$ and $f_{s}/f_{d}$.

These results are compatible with the current experimental averages~\cite{bib:hfagbase} and with available theoretical predictions~\cite{Li:2005kt,*Lu:2005mx,*DescotesGenon:2006wc,*Cheng:2009cn,*Williamson:2006hb,*Ali:2007ff,*Liu:2008rz,*Cheng:2009mu,*Mohanta:2000nk,*Kaur:2006yr,*Lu:2009cm}. 
The measurements of $\mathcal{B}\left(B^{0}_{s}\rightarrow K^{+}K^{-}\right)$, $\mathcal{B}\left(B^0_{s}\rightarrow\pi^+ K^-\right)$, $\mathcal{B}(B^0 \rightarrow K^+K^-)$ and $\mathcal{B}\left(\Lambda^0_b\rightarrow p\pi^-\right) /\, \mathcal{B}\left(\Lambda^0_b\rightarrow pK^-\right) $ are the most precise to date. Using a likelihood ratio test and including the systematic uncertainties on the signal yield, we obtain for the $B^0_s \rightarrow \pi^+\pi^-$ signal a significance of 5.3$\sigma$. This significance is estimated as $s_{\rm stat}=\sqrt{-2 \ln \frac{\mathcal{L}_{\rm B}}{\mathcal{L}_{\rm S+B}}}$, where $\mathcal{L}_{\rm S+B}$ and $\mathcal{L}_{\rm B}$ are the values of the likelihoods at the maximum in the two cases of signal-plus-background and background-only hypotheses, respectively. The value of $s_{\rm stat}=5.5\sigma$ is then corrected by taking into account the systematic uncertainty as $s_{\rm tot} = s_{\rm stat} / \sqrt{1 + \sigma_{\rm syst}^2/\sigma_{\rm stat}^2}$, where $\sigma_{\rm stat}$ and $\sigma_{\rm syst}$ are the statistical and systematic uncertainties. 
This is the first observation at more than $5\sigma$ of the $B^0_s \rightarrow \pi^+\pi^-$ decay.

\section*{Acknowledgements}

\noindent We express our gratitude to our colleagues in the CERN accelerator
departments for the excellent performance of the LHC. We thank the
technical and administrative staff at CERN and at the LHCb institutes,
and acknowledge support from the National Agencies: CAPES, CNPq,
FAPERJ and FINEP (Brazil); CERN; NSFC (China); CNRS/IN2P3 (France);
BMBF, DFG, HGF and MPG (Germany); SFI (Ireland); INFN (Italy); FOM and
NWO (The Netherlands); SCSR (Poland); ANCS (Romania); MinES of Russia and
Rosatom (Russia); MICINN, XuntaGal and GENCAT (Spain); SNSF and SER
(Switzerland); NAS Ukraine (Ukraine); STFC (United Kingdom); NSF
(USA). We also acknowledge the support received from the ERC under FP7
and the Region Auvergne.

\bibliographystyle{LHCb}
\bibliography{main}

\end{document}